\pdfoutput=1

\documentclass[preprints,article,accept,moreauthors,pdftex]{Definitions/mdpi}

\usepackage{tablefootnote}
\firstpage{1}
\pubvolume{xx}
\issuenum{1}
\articlenumber{5}
\pubyear{2020}
\copyrightyear{2020}
\history{Received: date; Accepted: date; Published: date}

\Title{Cosmogenic activation in double beta decay experiments}

\Author{Susana Cebrián $^{1,}$$^{2,*}$\orcidA{},}

\AuthorNames{Susana Cebrián}

\address{%
$^{1}$ \quad Centro de Astropartículas y Física de Altas Energías (CAPA), Universidad de Zaragoza, C/ Pedro Cerbuna 12, 50009 Zaragoza, Spain \\ 
$^{2}$ \quad Laboratorio Subterráneo de Canfranc, Paseo de los Ayerbe s/n, 22880 Canfranc Estación, Huesca, Spain}

\corres{Correspondence: scebrian@unizar.es}

\abstract{Double beta decay is a very rare nuclear process and, therefore, experiments intended to detect it must be operated deep underground and in ultra-low background conditions. Long-lived radioisotopes produced by the previous exposure of materials to cosmic rays on the Earth's surface or even underground can become problematic for the required sensitivity. Here, the studies developed to quantify and reduce the activation yields in detectors and materials used in the set-up of these experiments will be reviewed, considering target materials like germanium, tellurium and xenon together with other ones commonly used like copper, lead, stainless steel or argon. Calculations following very different approaches and measurements from irradiation experiments using beams or directly cosmic rays will be considered for relevant radioisotopes. The effect of cosmogenic activation in present and future double beta decay projects based on different types of detectors will be analyzed too.}

\keyword{neutrino; double beta decay; cosmic rays; activation; radioactive background}

\begin{document}



\section{Introduction}

There are still many open questions regarding the neutrino properties, even after the confirmation of their non-zero mass from oscillation experiments  \cite{nupdg}. The identification of a non-standard version of the double beta nuclear decay (DBD), without the emission of antineutrinos, could shed light on many of these questions: the determination of the Majorana or Dirac nature of neutrinos (whether or not neutrinos and antineutrinos are the same particle), the scale and ordering of the neutrino masses of the three generations, CP phases and the violation of the conservation of the lepton number. Consequently, the observation of this neutrinoless DBD would be very relevant for Nuclear and Particle Physics, Astrophysics and Cosmology. An overview of the investigation of DBD is made in  \cite{libroklapdor} and detailed reviews on the topic \cite{reviewfaessler,suhonen98,reviewelliot,reviewejiri,reviewavignone,vergados,reviewgiuliani,reviewgomez,reviewcremonesi,revieworo,reviewdolinski,appec}.

Double beta decay is a second-order standard weak process that some nuclei can undergo changing into an isobar. Two electrons and two antineutrinos are emitted in a double beta minus disintegration. The decay conserves the lepton number and is detectable for nuclei having an even number of neutrons and protons and the beta transition energetically forbidden or strongly suppressed by the angular momentum change. The DBD with emission of neutrinos, specifically reviewed in \cite{reviewsaakyan}, has been observed for several nuclei ($^{48}$Ca, $^{76}$Ge, $^{82}$Se, $^{96}$Zr, $^{100}$Mo, $^{116}$Cd, $^{128}$Te, $^{130}$Te, $^{136}$Xe, $^{150}$Nd and $^{238}$U) with measured half-lives varying approximately from $10^{19}$ to $10^{24}$~y (see recommended values at \cite{barabash19}).

A process without the emission of antineutrinos was proposed together with the standard (although rare) two-neutrino channel, accepting the violation of the lepton number conservation. Neutrinoless DBD has not evidenced for the moment, being the limits for the half-lives at the level of 10$^{25}$–10$^{26}$~y for double beta minus decays. Considering this type of DBD, the energy spectrum of the emitted electrons, which is relevant for detection, is different for the two DBD channels: in the neutrinoless decay, the two electrons take the transition energy, $Q$, and the spectrum of the sum energy is a peak at $Q$, while when the antineutrinos are emitted, electrons have a continuous energy spectrum. Another non-standard channel has been proposed too, with the emission of neutral bosons, like a Majoron $M$ (a Goldstone boson coupled to the exchanged virtual neutrino); the energy spectrum of electrons is continuous as for the two-neutrino channel but with different shape. In addition, as the beta decay can happen through the emission of electrons ($\beta^{-}$), positrons ($\beta^{+}$) or electron capture ($EC$), DBD can proceed by diminishing by two units the nuclear charge and these processes are also investigated \cite{reviewecec}. Transitions of DBD processes to excited states of the daughter nuclei can occur if kinematically allowed and are being studied too \cite{barabashexc}.

The transition probability for the neutrinoless DBD, being inversely proportional to the half-life $ T_{1/2}$, can be written (when considering only the mass term) as:
\begin{equation}
    (T_{1/2}^{0\nu})^{-1}=G_{0\nu}|M^{0\nu}|^{2} m_{\beta\beta}^{2},  \label{t0nu}
\end{equation}
with $G_{0\nu}$ the phase space integral and  $M^{0\nu}$ the nuclear matrix element. The neutrino effective mass is:
\begin{equation}
    m_{\beta\beta}=|\sum_{j=1}^{3} U_{ej}^{2} m_{j}|
\end{equation}
where $m_{j}$ is the mass of state $j$ and $U_{ej}$ are the elements of the mixing matrix between neutrinos $j$ and the electronic neutrino.

A lot of effort is being devoted for years to the identification of DBD due to its relevance. Different approaches have been followed, including the so-called  geochemical experiments (looking for an extra concentration of the daughter nuclei) or radiochemical experiments (trying to accumulate, extract and count the radioactive daughters of DBD emitters). But present searches are focused on direct counting experiments, registering the emitted electrons (energy spectrum and, in some cases, even tracks). Many different types of detectors, as germanium semiconductors, scintillators, gas chambers and bolometers, have been considered for DBD experiments, as discussed in specialized reviews \cite{reviewelliot,reviewgiuliani,reviewgomez,reviewcremonesi}.

Counting experiments require very special detection systems. Due to the very low probability of DBD, long data taking of a few years is mandatory and a large amount of DBD emitters has to be collected, being experiments at the tonne scale already considered. An excellent stability of the operation parameters is required during the whole data taking. Isotopic enrichment is necessary for most of the projects. In addition, good energy resolution is recommended to better single out the neutrinoless DBD peak. Having the DBD emitters inside the detector (detector$=$source approach) maximizes the signal detection efficiency, provided the detector is large enough to register all the energy released in the decay. Concerning the type of emitters, two properties must be analyzed: the transition energy $Q$ and the nuclear matrix element presented in  Eq.~\ref{t0nu} related to the transition probability. Since the radioactive background decreases with energy, isotopes with high $Q$ enjoy lower background levels in the region of the neutrinoless DBD signal

A particular challenge is common to all DBD experiments, as to other rare event searches: to minimize the background from different components in order to improve the signal sensitivity. Operation in deep underground sites sheltered from cosmic rays is a must. But even so, the material radioactivity (either in bulk or also on surfaces) from primordial, anthropogenic or cosmogenic nuclides, air-borne radon, radiogenic neutrons (from fission or ($\alpha$,n) reactions) or muons can generate background events. To take this under control, different strategies are implemented \cite{heusser,formaggio}: passive shieldings made of different materials (heavy ones like lead to attenuate the gamma background and light ones like water or polyethylene to moderate neutrons); veto systems to tag coincidence events and actively reject backgrounds; control and assessment of the material radiopurity for all the components used; and analysis methods to disentangle the DBD signal profiting from the particular features of each detector technology.

Long-lived radioisotopes induced by cosmic rays for the material exposure on the Earth's surface are a hazard for all experiments demanding ultra-low background conditions. The production by cosmic nucleons has been analyzed for many detector media (like germanium, sodium iodide, tellurium compounds, argon and xenon) as well as for other materials typically used in the experimental set-ups (like copper or stainless steel) in the context of the investigation of rare phenomena like the DBD and the dark matter direct detection, as reviewed in \cite{cebrianlrt2013,kudrylrt2017,cebriancosmogenic}. As other background sources are kept under control, the cosmogenic backgrounds are becoming increasingly relevant. The quantification of the yields of cosmogenic activation has usually important uncertainties as direct, experimental information is scarce.
The aim of this work is to review the studies developed to quantify and reduce the cosmogenic activation of materials performed in the context of DBD experiments. The structure of the paper is as follows. Section \ref{secact} sets up the basics of cosmogenic activation, describing the relevant processes, different approaches to estimate production cross sections and main components of the cosmic ray flux. Germanium, tellurium oxide and xenon are used in some of the largest DBD detectors investigating the neutrinoless DBD of $^{76}$Ge, $^{130}$Te and $^{136}$Xe; the studies for cosmogenic activation in these materials are reviewed in Secs. \ref{secge}, \ref{secte} and \ref{secxe}, respectively, while Sec. \ref{secotherDBD} presents studies performed for materials used in experiments focused on other DBD emitters like $^{82}$Se, $^{100}$Mo or $^{150}$Nd. For each material, the status of the corresponding DBD experiments will be briefly summarized, the principal cosmogenic products will be presented, the activation works (based on measurements or different calculations) from nucleons and muons, if available, will be described and the effect of cosmogenics in the background levels will be discussed. Whenever possible, production rates of the induced isotopes at sea level are summarized in tables for the different targets, as this is the key element to evaluate cosmogenic activity in particular circumstances. Similarly, Sec. \ref{othernonDBD} presents the activation studies for other materials typically used in DBD experiments like copper, stainless steel or argon. Finally, a summary and conclusions are given in  Sec. \ref{secsum}.

\section{Cosmogenic activation}
\label{secact}

Radioactive impurities in the components used in a low-background experiment induced by the exposure to cosmic rays at the surface (during production, transportation or storage) may become very problematic, being in some cases even more relevant than the primordial activity. Indeed, the limited knowledge of cosmic ray activation was considered in~\cite{gondolo} as one of the three ``main uncertain nuclear physics aspects of relevance in the direct detection of dark matter''. Production of cosmogenic radioactive isotopes is also considered in other fields \cite{lal,beer} as some products are relevant in  different contexts related to Astrophysics, Geophysics or Archaeology.

The spallation of nuclei by high energy nucleons is one of the dominant processes for the cosmogenic production of radionuclides, but other reactions like fragmentation, fission, break-up or capture are important too. Spallation reactions produce the emission of neutrons and charged particles together with the generation of residual nuclei far from the target. They are typically described in two steps: a first one leading to an excited remnant nucleus and a second one, much slower, corresponding to the de-excitation of this nucleus. Isotope production at the Earth's surface is dominated by neutrons because protons are absorbed by the atmosphere. But at high altitudes protons are also relevant for activation and the cosmic flux increases.

Cosmogenic activation of materials underground can be considered in many cases negligible, as the flux of cosmic nucleons is suppressed just for depths of a few tens of meter water equivalent (m.w.e.) and the neutron fluxes in deep underground facilities are orders of magnitude lower than at surface. Radiogenic neutrons have energies (around a few MeV) too loo to produce spallation processes. Therefore, activation underground is mainly due to muons. Negative muon capture is dominant at shallow depths while deep underground fast muon interactions are the most relevant ones: muon spallation (virtual photon nuclear disintegration) and electromagnetic and nuclear reactions from secondary particles. As the muon energy spectra and fluxes depend on depth, underground activation can be very different for different sites. Studies were made for neutrino detectors using large amounts of liquid scintillator; but muon activation has been assessed for some DBD detectors too, as it will be shown. This on-site activation can be problematic for next generation experiments and may set a minimum for the required depths.

The relevant cosmogenic radioisotopes generated depend on the  target material, but some spallation products like tritium are commonly induced. Tritium is a pure $\beta^-$ emitter with $Q=(18.591\pm0.001)$~keV and a half-life of $(12.312\pm0.025)$~y \cite{ddep}; therefore, it is very relevant for dark matter searches when produced in the detector medium \cite{tritiumpaper}, but has no effect for most of the DBD searches. Estimates of tritium yields require the identification of all the final reaction products.

Cosmogenic activation can be minimized by reducing the surface exposure, using shieldings against the cosmic rays, avoiding flights and storing, or even producing, materials underground. Purification techniques can eliminate also many of the induced  isotopes. But these preventive measures make the experiment preparation more complex, for instance for crystal growth or detector mounting. Consequently, it would be advisable to assess the relevance of the material exposure to cosmic rays for the experiments and its effect on the sensitivity. To quantify the induced activity, $A$, of an isotope with decay constant $\lambda$, both  the production rate $R$ of the isotope in the considered target as well as the exposure history must be well-known. In particular, $A$ can be computed as:
\begin{equation}
A = R [1-\exp(-\lambda t_{exp})] \exp(-\lambda t_{cool}),
\end{equation}
\noindent considering $t_{exp}$ the time of exposure to cosmic rays and $t_{cool}$ the cooling time (time spent underground once shielded from cosmic rays).

Some direct measurements of productions rates have been carried out for a few materials from the saturation activity, obtained by sensitive screening of materials exposed in well-controlled conditions. But in many cases production rates must be evaluated from the flux (per unit energy) of cosmic rays, $\phi$, and the isotope production cross-section, $\sigma$, being both ingredients dependent on the particle energy $E$:
\begin{equation}
R=N_t\int\sigma(E)\phi(E)dE \label{eqrate},
\end{equation}
\noindent with $N_t$ the number of target nuclei.

\subsection{Production cross sections}

Excitation functions for the production by nucleons of a particular nuclide in a target over a wide range of energies (from a few MeV up to tens or hundreds of GeV) cannot be obtained just from measurements of production cross-sections with beams. Calculations must be made to have complete information on the excitation functions. The EXFOR database (CSISRS in USA) \cite{exfor} compiles nuclear reaction data and then cross sections for a particular target, projectile, energy or reaction can be searched for. Experimental data on production cross sections are scarce but essential to validate  calculations.

Different approaches can be considered to compute production cross sections. Semiempirical formulae have been derived for nucleon-nucleus reactions. The Silberberg \& Tsao equations presented in Refs. \cite{tsao1,tsao2,tsao3} can be used for targets with mass number A$\geq$3, for products with A$\geq$6 and for energies $>$100~MeV. They have been integrated in different codes: COSMO (FORTRAN program) \cite{cosmo}, YIELDX (FORTRAN routine, including the latest updates of the equations) \cite{tsao3} and, more recently, ACTIVIA (C++ computer package, using also experimental data when available) \cite{activia}. Any of these codes offers very fast calculations, although as the formulae are based only on proton-induced reactions, neutrons and protons cross sections must be implicitly assumed to be equal.

The other approach in production cross sections calculations is the Monte Carlo (MC) simulation of the  interaction between nucleons and nuclei. Modeling properly the interactions for isotope production requires the consideration of different reactions: from the formation and decay of compound nuclei to the intranuclear cascade of nucleon interactions followed by de-excitation processes like fission, fragmentation, spallation or breakup. As described in detail in Ref. \cite{david}, many different models and codes have been developed and validated to implement this in different contexts, like for instance the production of medical radioisotopes, the transmutation of nuclear waste, the prevention of damage to electronics on spacecraft, the radioprotection of astronauts, or studies of comic rays and astrophysics. Some of these codes have been implemented in general-purpose codes like GEANT4 \cite{geant4}, FLUKA \cite{fluka} and MCNP \cite{mcnp}. Evaluated libraries of production cross sections have been elaborated, providing different coverage of reactions, projectiles and energies, like for example TENDL (TALYS-based Evaluated Nuclear Data Library) \cite{tendl} (based on the TALYS code, for protons and neutrons with energies up to 200~MeV) or HEAD-2009 (High Energy Activation Data) \cite{head2009} (for protons and neutrons with higher energies, from 150~MeV up to 1~GeV).

\begin{figure}
 \includegraphics[width=0.9\textwidth]{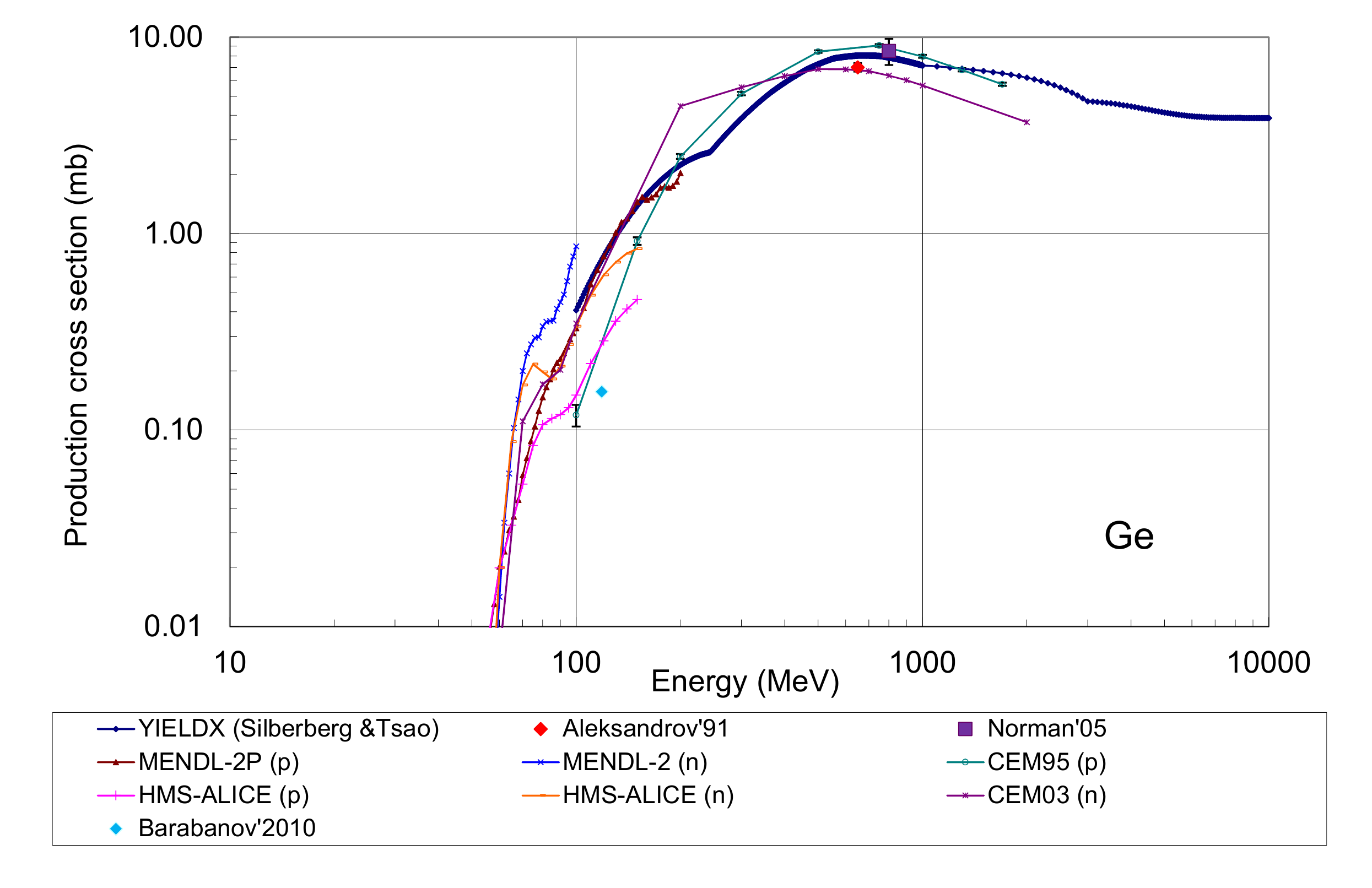}
  \includegraphics[width=0.9\textwidth]{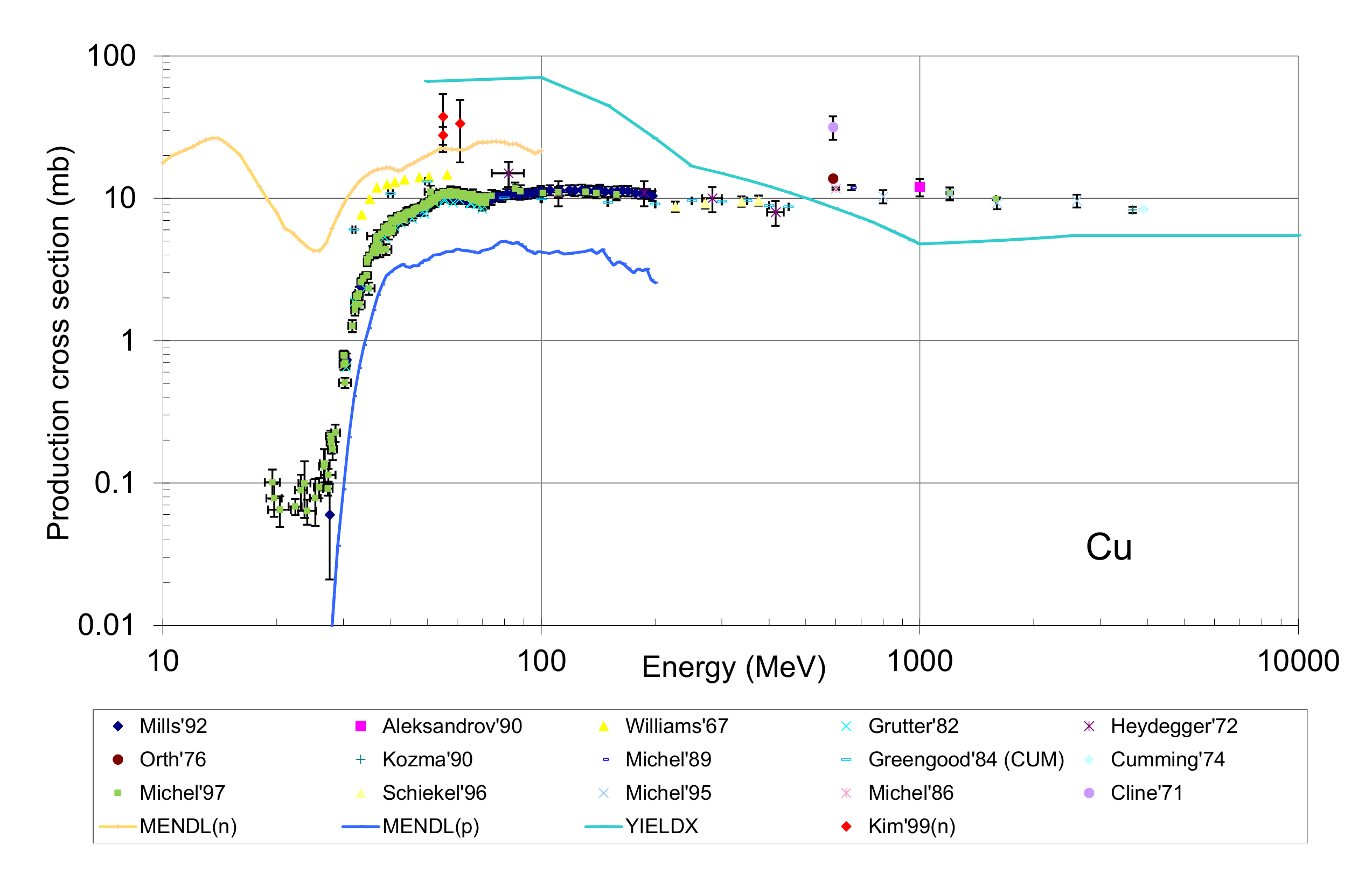}
 \caption{Compilation of excitation functions for the production of $^{60}$Co by protons and neutrons in natural germanium (top) and in copper (bottom). Experimental data obtained from the EXFOR database are shown together with calculations using semiempirical formulae and MC simulations.}
 \label{excco60}
\end{figure}

A more detailed discussion on the different options to undertake the evaluation of cosmogenic production cross sections and yields can be found in \cite{cebriancosmogenic}. There is no perfect  approach, depending the suitability of a model or code to a particular situation on targets,  projectiles and energies. In \cite{cebriancosmogenic}, to analyze a particular product and target, it is proposed to collect and analyze firstly all the information on the excitation functions by neutrons and protons from all the available sources (calculations and experiments); and then, to choose the best description by minimizing deviations. Systematic comparisons between measurements and calculations are usually made based deviation factors, $F$, defined as:
\begin{equation}
F=10^{\sqrt{d}},\hspace{0.5cm}
d=\frac{1}{n}\sum_{i}(\log\sigma_{exp,i}-\log\sigma_{cal,i})^{2},
\label{eqF}
\end{equation}
\noindent with $n$ the number of pairs of measured  and calculated cross sections $\sigma_{exp,i}$ and $\sigma_{cal,i}$ at a certain energy. As an example, Fig. \ref{excco60} shows a compilation of excitation functions for $^{60}$Co, a typical cosmogenic product in different targets, generated in natural germanium and in copper by nucleons taken from different sources, as made in Ref. \cite{cebrian}; the availability of a large amount of experimental data in copper allows a more reliable validation of calculations.

\subsection{Cosmic ray flux}
\label{secflux}

Together with the production cross sections, the other element to calculate the production rates from Eq.~\ref{eqrate} is the cosmic ray flux and its energy distribution. Nucleons, with energies from MeV to GeV, are responsible of most of the activation produced on the surface. At sea level, the the number of protons and neutrons with energies of a few GeV is roughly the same; but at lower energies, as charged particles are absorbed in the atmosphere, the proton to neutron ratio is much lower than one, being of only 0.03 at 100~MeV \cite{lal}. This is the reason why neutrons produced the bulk of the cosmogenic activation at sea level. But proton activation is not completely negligible; its contribution is quoted in \cite{lal} as $\sim$10\%, in agreement with results for proton activation in germanium \cite{barabanov,wei}. Activation from other particles like muons can be even smaller \cite{mei2016,wei}.

\begin{figure}
 \includegraphics[width=\textwidth]{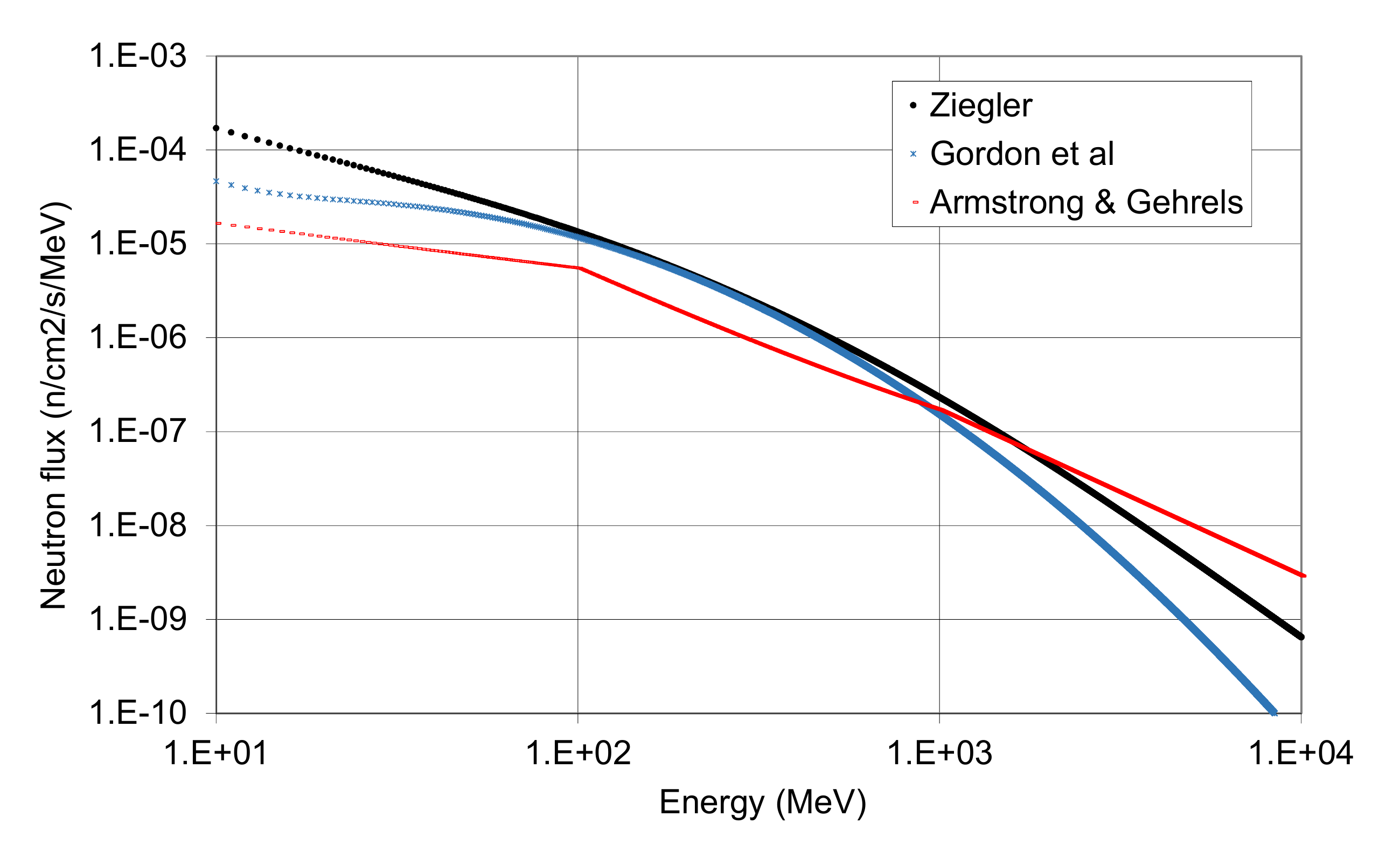}
 \caption{Differential neutron flux as derived from parameterizations at sea level by Armstrong and Gehrels \cite{armstrong,gehrels}, Ziegler \cite{ziegler} and Gordon et al. \cite{gordon}.}
 \label{spc}
\end{figure}

Different descriptions of the energy spectrum of neutrons at sea level have been considered for cosmogenic activation studies, like the ones presented at Refs. \cite{hess} and \cite{lal}. Parameterizations given in~\cite{armstrong,gehrels} are used in the ACTIVIA code. In~\cite{ziegler}, following a revision of all previous available results, a new parameterization valid for energies from 10~MeV to 10~GeV was provided. After a campaign of cosmic neutron measurements in the US, a different analytic function from the fit to data for energies greater than 0.4~MeV was proposed by Gordon et al. \cite{gordon}. Three different parameterizations (corresponding to the flux at sea level in New York City) are compared in Fig.~\ref{spc}; in the range 10~MeV-10~GeV, the integral flux is 5.6 (3.6) $\times 10^{-3}$cm$^{-2}$s$^{-1}$ from \cite{ziegler} (\cite{gordon}). It is worth noting that to assess activation at a certain location, the evaluated flux for New York City must be properly scaled using available factors \cite{ziegler} or tools \cite{calculator}.

For the energy spectrum of surface protons the CRY (``Cosmic-ray Shower Library'') generator \cite{hagmann} can be used. At sea level, the integral flux in 100~MeV-100~GeV is 1.36$\times 10^{-4}$cm$^{-2}$s$^{-1}$ while the energy spectra  for different particles (like nucleons and muons) is obtained from the full MC simulation of primary protons and atmospheric showers. The good agreement with data shows that these calculations give reliable description of the distributions of cosmic rays at sea level. As discussed in Ref. \cite{kudrylrt2017}, CRY allows the generation of  energies, positions and directions of different  particles; but primary nuclei are not simulated (only protons are considered) and the accuracy of the derived spectra is limited (due to the energy bins defined). The flux of protons and high-energy gamma rays can be also obtained from the EXPACS (``EXcel-based Program for calculating Atmospheric Cosmic-ray Spectrum'') sofware program \cite{expacs}. It calculates terrestrial fluxes of nucleons, ions with charge below 28, muons and other particles for different positions and times in the Earth's atmosphere. Based on these calculated fluxes, EXPACS can also deduce the effective dose, ambient dose equivalent, and absorbed dose in air due to cosmic-ray exposure.

It is also possible to accomplish the full calculation described in Eq. \ref{eqrate} with one code. The CONUS (``COsmogenic NUclides Simulations'') toolkit, developed for the analysis of cosmogenic radioisotopes in extra-terrestrial and terrestrial objects \cite{conus}, can be used to compute production rates in detector materials too. It is based on the MCNPX and MCNP codes combined with LAHET (code for transport of nucleons, pions, muons and light nuclei). An isotropic irradiation with primary galactic cosmic-ray protons is used in calculations. Measurements and TALYS calculations are used for cross sections. Input particles in  CONUS are considered with energies from 0.001~eV to 20~GeV.

\section{Germanium}
\label{secge}

Germanium experiments play a leading role in searches for DBD. They have derived very important results and offer now very good prospects. Activation in germanium has been extensively analyzed, much more than in any other DBD target. A brief overview of germanium experiments for DBD is presented here, before the detailed description of activation studies in this material.

\subsection{Germanium DBD experiments}

The use of germanium detectors to investigate the DBD of $^{76}$Ge started over five decades ago \cite{morales03}. They have excellent energy resolution and radiopurity. In particular, for DBD investigation, being based on the detector$=$source approach, the signal detection efficiency is high, the nuclear matrix element is favorable and the transition energy $(Q=(2039.061\pm0.007)$~keV \cite{qge}) is reasonably high. Effective pulse shape discrimination techniques have been implemented: DBD events deposit energy at one site while most of the backgrounds generate multi-site deposits producing distinguishable electronic signals \cite{igexpsd,psdradial1,klapdortracks2}. On the other hand, the natural isotopic abundance of $^{76}$Ge is low (7.8\% \cite{reviewsaakyan}), which makes enrichment necessary.

The study of the neutrinoless DBD of $^{76}$Ge using enriched semi-coaxial HPGe detectors provided the strongest bounds to the effective neutrino mass in the nineties, from IGEX (``International Germanium EXperiment'') operated in the Canfranc Underground Laboratory (LSC) in Spain \cite{igexfinal,lscppnp} and the Heidelberg-Moscow (HM) experiment \cite{GUN97,hm}, at the Gran Sasso Underground Laboratory (LNGS) in Italy. HM presented 90 \% C.L. limits for the half-life of $^{76}$Ge and the effective mass ($T_{1/2}^{0\nu}\geq 1.9 \times 10^{25}$~y and $m_{\beta\beta} \leq 0.35$~eV \cite{klapdor01}), but a few members of the collaboration later claimed hints of a positive signal \cite{klapdorevidence4}, with a 4.2$\sigma$ confidence level, corresponding to  $T_{1/2}^{0\nu}=1.2 \times 10^{25}$~y and $m_{\beta\beta}=0.44$~eV. New strategies based on  new detector designs (like Broad Energy germanium (BEGe) detectors \cite{broadenergy1}) were developed to further suppress the radioactive background of germanium experiments; discrimination techniques following different approaches are under consideration \cite{psdgerda2019,psdmajorana2019}.

The ``GERmanium Detector Array'' (GERDA) \cite{gerdascience} operates at LNGS with enriched germanium detectors in cryogenic liquid to minimize the surrounding materials. In the Phase I of GERDA \cite{gerdaphaseI}, the enriched detectors used by IGEX and HM (with a total mass of $\sim$18~kg) were refurbished for bare operation in liquid argon. A precise estimate of the half-life of the two-neutrino DBD of $^{76}$Ge was derived as $T_{1/2}^{2\nu}=(1.926 \pm 0.094)\times 10^{21}$~y, together with new limits for other DBD modes \cite{gerda2nuM,gerdaexcited,gerdaECAr}. For the Phase II of GERDA, since 2015, 30~BEGe detectors were added \cite{gerdabe}. The improved performance of the detectors and of the liquid argon active veto system \cite{gerdaIIupgrade,gerdanature}, has allowed to decrease the background level down to 5.2$\times$10$^{-4}$~counts keV$^{-1}$ kg$^{-1}$ y$^{-1}$ \cite{gerdafinal}. From the final Phase I and II data, corresponding to 127.2~kg$\cdot$y, the 90\% C.L. limit for the half-life is $T_{1/2}^{0\nu}>1.8\times 10^{26}$~y; this gives a limit for the effective neutrino mass of 79-180~meV \cite{gerdafinal}.

The \textsc{Majorana Demonstrator}, built and operated by the MAJORANA Collaboration, consists of an array of ultra-low background HPGe detectors with a total mass of 44~kg (29.7~kg enriched in $^{76}$Ge) installed in the Sanford Underground Research Facility (SURF) in Lead, South Dakota, US, mainly for DBD investigation. A careful processing procedure to minimize cosmogenic activation of detectors was followed in the fabrication of the point-contact Ge detectors using germanium isotopically enriched to 88\% in $^{76}$Ge \cite{majprocessing}. The data taking with the entire array of detectors started in 2016 and the measured background level is $(11.9\pm2.0)$ counts/(FWHM t y) \cite{maj2019}. From the latest results presented, the 90\% C.L. limit for the neutrinoless DBD half-life of $^{76}$Ge is $T_{1/2}^{0\nu}>2.7 \times 10^{25}$~y and for the effective neutrino mass the range 200-433~meV is derived depending on the considered matrix elements. Results for DBD to excited states \cite{majexc} and  for other rare events have been also obtained  \cite{majbosonic,maje,majnuc}.

The LEGEND (``Large Enriched Germanium Experiment for Neutrinoless $\beta\beta$ Decay'') collaboration
has been created following the success of GERDA and the \textsc{Majorana Demonstrator}. The goal is to implement a tonne-scale experiment for $^{76}$Ge profiting from the superior energy resolution and background discrimination capabilities of germanium detectors, to explore after ten years of data taking a neutrinoless DBD half-life beyond 10$^{28}$~y and the full inverted hierarchy region of neutrino masses \cite{appec,legend}. The plan is to start with 200~kg using the existing GERDA facilities at LNGS; the LEGEND program is supported by the Double Beta Decay APPEC Committee \cite{appec}.

\subsection{Activation studies}

Together with DBD experiments, other projects investigating rare events, like the galactic dark matter interaction, are using germanium crystals either as pure ionization detectors (like CoGENT \cite{cogent}, TEXONO or CDEX \cite{cdex}) or as cryogenic detectors  measuring simultaneously ionization and heat (like CDMS \cite{supercdms} or EDELWEISS \cite{edelweiss}). For this reason, a large number of studies have been carried out for activation in germanium. For neutrinoless DBD searches, only $^{60}$Co (T$_{1/2}$=5.27 y, $\beta^{-}$ emitter with transition energy of 2823.1 keV and gamma emissions of 1173.3 and 1332.5 keV) and $^{68}$Ge (T$_{1/2}$=270.8 d, decaying by electron capture to $^{68}$Ga, $\beta^{+}$/EC emitter with transition energy of 2921.1 keV) are relevant, due to the continuous spectrum beta emission at the region of interest of 2039 keV. But many other isotopes have been studied too, like $^{54}$Mn, $^{55}$Fe, $^{56}$Co, $^{57}$Co,  $^{58}$Co, $^{63}$Ni or $^{65}$Zn. Production of $^{68}$Ge is relevant also in other contexts, like  medical diagnosis due to Positron Emission Tomography. The production of tritium in germanium is particularly worrisome for dark matter experiments; indeed, it was proposed as a possible explanation of the low energy background measured in the IGEX detectors \cite{cebriantaup2003} and tritium has been highlighted as a significant background for SuperCDMS \cite{supercdms} and CDEX \cite{cdexcos}.

In the rest of this section, the main results related to the cosmogenic activation in enriched and, for completeness also in natural germanium, will be described; the enriched germanium used in DBD experiments has typically an isotopic composition of 86\% of $^{76}$Ge and 14\% of $^{74}$Ge. Tables~\ref{geenr} and \ref{genat} summarize the production rates at sea level obtained in the different estimates for both germanium compositions.

\begin{table}
\caption{Rates of production of long-lived radioisotopes induced in enriched germanium (as used in DBD experiments, typically 86\% of $^{76}$Ge and 14\% of $^{74}$Ge) at sea level. Measurements from  Refs.~\cite{heusser,elliot2010,majoranalrt} and different calculations are presented (see text). All numbers are expressed in atoms/kg/day.}
\centering
\tablesize{\footnotesize}
\begin{tabular}{@{}lcccccccc@{}}
\toprule
&  \textbf{$^{3}$H} &  \textbf{$^{54}$Mn} & \textbf{$^{55}$Fe} & \textbf{$^{57}$Co} & \textbf{$^{58}$Co} & \textbf{$^{60}$Co} & \textbf{$^{65}$Zn} & \textbf{$^{68}$Ge} \\
\midrule
Half-life \cite{ddep} & 12.312(25) & 312.19(3) & 2.747(8) & 271.81(4) & 70.85(3) & 5.2711(8) & 244.01(9) & 270.95(26)  \\
units & y & d & y & d & d & y & d & d \\ \hline
Measurement \cite{heusser} && 2.3 & &1.6 & 1.2 & & 11 & \\
Measurement \cite{elliot2010} && 2.0$\pm$1.0 & & 0.7$\pm$0.4 &  & 2.5$\pm$1.2 & 8.9$\pm$2.5 & 2.1$\pm$0.4 \\
Meas. (MAJORANA) \cite{majoranalrt} & 140$\pm$10 & 4.4$\pm$4.1 & 2.1$\pm$0.7 & & & & 4.3$\pm$3.6 & 3.3$\pm$1.6 \\
Monte Carlo \cite{avignone} & 140 & 1.4 && 1 &1.8 &   &  6.4 & 0.94  \\
Monte Carlo \cite{miley} &&& & 0.08 & 1.6 &  3.5 & 6.0 & 1.2 \\
SHIELD \cite{barabanov}  &&&  &  && 3.3 & & 5.8 \\
TALYS \cite{mei} & 24.0 & 0.87 & 3.4 & 6.7 &  & 1.6  & 20 & 7.2 \\
MENDL+YIELDX \cite{cebrian} && 3.7 & 1.6 & 1.7 & 4.6 & 5.1 & 20 & 12 \\
TENDL+HEAD\cite{tritiumpaper} & 94$\pm$34 & & & & & & & \\
ACTIVIA \cite{activia} && 2.2 & 1.6 & 2.9 & 5.5 & 2.4 & 10.4 & 7.6 \\
ACTIVIA \cite{wei} & 51.3 & 2.2 & 1.2 & 2.3 & 5.5 & 4.4 & 9.7 & 15.4 \\
GEANT4 \cite{wei} & 47.4 & 1.4 & 4.5 & 3.3 & 2.9 & 2.4 & 24.9 & 21.8 \\
GEANT4+CRY \cite{cdexcos} & 22.8 & 0.96 & 2.9 & 2.8 & & 1.9 & 18.0 & 20.0 \\ 
\bottomrule
\end{tabular} \label{geenr}
\end{table}

There are many studies based on the MC simulation of the nucleon interaction in the target following different models.
\begin{itemize}
\item First estimates of production rates for germanium were presented in Refs.~\cite{avignone,miley} from excitation functions computed with the spallation codes LAHET/ISABEL and the neutron spectra from \cite{hess,lal}, including a calculation for tritium; values shown in Tables~\ref{geenr} and \ref{genat} correspond to those using the Hess spectrum \cite{hess}. Experimental estimates of the production rates were also derived from IGEX data taken in Canfranc and Homestake laboratories with exposed germanium detectors \cite{avignone}. Agreement between calculations and measurements was found to be within a factor of 2.
\item The SHIELD code was used for the excitation functions considered in the production rates of $^{60}$Co and $^{68}$Ge estimated in Ref.~\cite{barabanov}, including both neutron and proton contributions; as mentioned before in section~\ref{secflux}, protons produce $\sim$10\% of the total rate.
\item TALYS was used for the excitation functions considered in the estimate of production rates in Ref.~\cite{mei}. The Gordon et al. parameterization was assumed for the neutron spectrum.
\end{itemize}

Other activation studies are based on codes implementing semiempirical formulas for the production cross sections.
\begin{itemize}
\item A semiempirical code referred to as $\Sigma$ was used in the calculations in Ref.~\cite{genius}.
\item The ACTIVIA code, using the parameterization from \cite{armstrong,gehrels} for the neutron energy spectrum, was applied to deduce production rates for benchmark in both enriched and natural germanium \cite{activia}.
\end{itemize}

In other cases, both approaches (MC simulation and semiempirical formulas) are considered and also general purpose simulation packages have been used.
\begin{itemize}
\item In Ref.~\cite{cebrian}, production rates were calculated after a careful collection and evaluation of excitation functions from experimental data and different calculations (using YIELDX or from MENDL \cite{mendl} and other libraries \cite{study}). The computed deviation factors allowed the best selection for the cross sections: from HMS-ALICE and from YIELDX below and above 150~MeV, respectively. Neutron cosmic spectra from both \cite{ziegler} and \cite{gordon} were considered, finding that estimates using the Gordon et al. spectrum were usually closer to experimental results.
\item Cosmogenic activation for both natural and enriched germanium was studied in detail in Ref.~\cite{wei}, calculating many different  production rates from neutrons, protons and muons, using GEANT4 and ACTIVIA (results from this work presented in Tables~\ref{geenr} and \ref{genat} correspond to neutrons and the Gordon et al. spectrum). Moreover, the expected counting rates from activation were assessed assuming certain  exposure and cooling conditions and the effectiveness of shieldings for activation was analyzed too.
\item In Ref.~\cite{cdexcos}, another study considering natural and $^{70}$Ge depleted germanium detectors (with 86.6\% of $^{76}$Ge) was carried out using in this case GEANT4 and the CRY library to generate particle showers including nucleons, muons and others. Production rates were evaluated and compared for different locations and altitudes and the results validated against CDEX-1B detector data.
\item The same approach based on GEANT4 and CRY considering neutrons, protons, muons and gammas has been followed to optimize the design of a shielding for the transport and storage of high-purity germanium \cite{yan2020}. Six materials (iron, copper, lead, liquid nitrogen, polyethylene, and concrete) have been considered. Relevant production rates have been calculated for natural
germanium at sea level with and without shield. Iron is confirmed to be an optimal shielding material, thanks to the lower production of secondary neutrons; but an optimized shielding structure using different materials is shown to be more effective, reducing by approximately one order of magnitude the production rates of the cosmogenic radioisotopes.
\item Calculations using GEANT4 and CONUS software packages and ACTIVIA (with the default input neutron spectrum) were used to quantify production rates and estimate the corresponding counting rates in the Obelix HPGe detector (with mass of 3.19 kg), operating in the Modane underground laboratory in France \cite{obelix}. Interactions of the secondary cosmic nucleons for several components, including the Ge crystal, were considered to quantify the production of several isotopes. In general, the underproduction observed with ACTIVIA can be due to the fact that the energy threshold in CONUS is lower than in ACTIVIA. The simulated gamma ray spectra obtained from the calculated production rates of cosmogenic radionuclides were in good agreement with measurements from the Obelix detector. Overall, it was found that the contribution of cosmogenics to the detector background decreased from 39\% (after 10 months of cooling down) to 14\% (after three years).
\item In Ref.~\cite{tritiumpaper}, production rates specifically for tritium were deduced for common target materials in dark matter detectors, including germanium, following a selection of excitation functions including mainly those from TENDL and HEAD-2009 libraries at low and high energies, respectively. The result for natural germanium is in very good agreement with the measured production rates by EDELWEISS \cite{edelweisscos} and CDMSlite \cite{cdmslite}. As it can be seen from Tables~\ref{geenr} and \ref{genat}, the measured production rate by MAJORANA for enriched germanium is higher than for natural germanium; the explanation could be that cross sections increase with the mass number of the germanium isotope, following TENDL-2013 and HEAD-2009 data \cite{tritiumpaper}.
\end{itemize}

Calculation of activation yields are necessary and very important, but validation with experimental data is essential. Some irradiation experiments with neutron beams have been made.
\begin{itemize}
\item A target made of natural germanium was irradiated at Los Alamos Neutron Science Center (LANSCE) with a 800~MeV proton beam and production cross sections were derived after the screening of the sample up to 5~years later with germanium detectors at Berkeley \cite{norman}. The results agree reasonably with the estimates from the Silberberg\&Tsao formulas.
\item In Ref.~\cite{elliot2010}, a sample of enriched germanium was irradiated also at LANSCE but with a neutron beam with energies up to $\sim$700~MeV resembling the cosmic neutron spectrum. Gamma counting with germanium detectors was carried out at the Waste Isolation Pilot Plant (WIPP) after cooling to quantify the production of radioisotopes. Cross sections were also calculated using the CEM03 code, finding in general overestimated values. Production rates were deduced from the measurements taken into considerations the the Gordon et al. neutron spectrum.
\item Neutron irradiation on enriched germanium allowed to measure cross section for the radiative capture of neutrons on $^{74}$Ge and $^{76}$Ge for thermal energies \cite{ncapture1,ncapture2} and for energies of a few MeV \cite{ncapture3}. These results are particularly relevant for DBD experiments. In addition to the de-excitation gamma emissions from the generated nucleus following the neutron capture, the nucleus can be radioactive and decay.
\end{itemize}

The analysis of data taken by germanium experiments, looking for dark matter or DBD, has allowed to quantify activation yields deriving very important results.
\begin{itemize}
\item A detailed analysis of background data taken by the EDELWEISS experiment for a long time and using different germanium detectors with a different, well-known exposure history to cosmic rays allowed to derive production rates of several radioisotopes induced in natural germanium, including, for the first time, tritium \cite{edelweisscos}. Low energy data were fitted considering a continuum and peaks at the electron binding energies for K, L shells, produced by induced nuclei decaying by EC. The obtained production rates were compared with ACTIVIA calculations using the Gordon et al. parameterization for the neutron spectrum.
\item Measurements of the production rates of tritium and other cosmogenic isotopes were carried out also with the CDMS low ionization threshold experiment (CDMSlite) \cite{cdmslite}, from the analysis of data from the second run. The measured spectrum was modeled and fit considering  contributions from different isotopes including tritium. Thanks to the knowledge of the well documented exposure history of the detector, production rates at sea level could be derived. In addition, estimates of the rates using TALYS and INCL++-ABLA codes below and above 100 MeV, respectively, were presented in Ref.~\cite{cdmslite}, using the Gordon et al. neutron spectrum and taking into account the contribution by protons too.
\item When processing enriched germanium detectors for the {\sc{Majorana Demonstrator}}, a lot of care was taken to minimize cosmogenic activation. As described in Ref.~\cite{majoranacrystals}, whenever possible the material was shielded or stored underground to suppress specially the production of $^{68}$Ge, as this isotope cannot be eliminated by zone refinement and crystal growth. It is estimated that the steel shield used during transportation from Russia to US reduced the $^{68}$Ge formation by a factor of ten. In the data of the {\sc{Majorana Demonstrator}}, a factor 30 reduction for the Ga X-ray peak (following the EC decay of $^{68}$Ge) has been measured in enriched crystals over that in natural germanium detectors not shielded; a similar reduction could be expected at higher energies for the emissions of $^{68}$Ge and $^{60}$Co \cite{majoranacrystals}. Tritium has been observed too in the detectors of the {\sc{Majorana Demonstrator}}, having a well-known exposure history. Production rates of different radionuclides in enriched germanium have been reported~\cite{majoranalrt}, assuming as in other cases for the fitting model a flat background, different X-ray peaks and the tritium beta spectrum.
\end{itemize}

\begin{table}
\caption{Rates of production of long-lived radioisotopes induced in natural germanium at sea level. Measurements from  Refs.~\cite{avignone,edelweisscos} and different calculations are presented (see text). All numbers are expressed in atoms/kg/day.}
\centering
\tablesize{\footnotesize}
\begin{tabular}{@{}lccccccccc@{}}
\toprule
& \textbf{$^{3}$H} & \textbf{$^{49}$V} & \textbf{$^{54}$Mn} & \textbf{$^{55}$Fe} & \textbf{$^{57}$Co} & \textbf{$^{58}$Co} & \textbf{$^{60}$Co} & \textbf{$^{65}$Zn} & \textbf{$^{68}$Ge} \\
\midrule
Half-life \cite{toi,ddep} & 12.312(25) & 330~d & 312.19(3) & 2.747(8) & 271.81(4) & 70.85(3) & 5.2711(8) & 244.01(9) & 270.95(26)  \\
units & y & d & y & d & d & y & d & d \\ \hline
Measurement \cite{avignone} & & &  3.3$\pm$0.8 & &  2.9$\pm$0.4 &  3.5$\pm$0.9 & & 38$\pm$6 & 30$\pm$7 \\
Meas. (EDELWEISS) \cite{edelweisscos} & 82$\pm$21  & 2.8$\pm$0.6 & & 4.6$\pm$0.7 & & & & 106$\pm$13 & $>$71 \\
Meas. (CDMSlite) \cite{cdmslite} & 74$\pm$9 & & & 1.5$\pm$0.7 & & & & 17$\pm$5 & 30$\pm$18 \\
Monte Carlo \cite{avignone} & 210  &  & 2.7 & & 4.4 &   5.3  &  &  34.4 &  29.6       \\
Monte Carlo \cite{miley} & &   & & & 0.5 & 4.4 &  4.8 & 30.0 & 26.5 \\
Sigma \cite{genius} & &   & 9.1 & 8.4 & 10.2 & 16.1 & 6.6 &79.0 & 58.4 \\ 
SHIELD \cite{barabanov} & &  & &  & & & 2.9  &  & 81.6  \\
TALYS \cite{mei} & 27.7 & & 2.7  & 8.6  & 13.5   & & 2.0 & 37.1  & 41.3 \\
TALYS+INCL++-ABLA \cite{cdmslite} & 95 & & & 5.6 & & & & 51 & 49 \\
MENDL+YIELDX \cite{cebrian} & &  & 5.2 &   6.0 & 7.6 & 10.9   & 3.9  & 63  & 60 \\
TENDL+HEAD\cite{tritiumpaper} & 75$\pm$26 & & & & & & & & \\
ACTIVIA \cite{activia} & & & 2.7 & 3.4 & 6.7 & 8.5 & 2.8 & 29.0 & 45.8 \\
ACTIVIA \cite{edelweisscos} & 46 & 1.9 & & 3.5 & & & & 38.7 & 23.1  \\
ACTIVIA (MENDL-2P) \cite{edelweisscos} & 43.5 & 1.9 & & 4.0 & & & & 65.8 & 45.0 \\
ACTIVIA \cite{wei} & 52.4 & & 2.8 & 4.1 & 8.9 & 11.4 & 4.1 & 44.2 & 24.7 \\
ACTIVIA \cite{obelix} & 30 & & 3 & & 6 & & 3 & 20 & 10 \\
GEANT4 \cite{wei} & 47.4 & & 2.0 & 7.9 & 7.4 & 5.7 & 2.9 & 75.9 & 182.8 \\
GEANT4+CRY \cite{cdexcos} & 23.7 & 1.4 & 0.94 & 4.2 & 4.7 & & 1.5 & 40.5 & 83.1 \\ 
GEANT4+CRY \cite{yan2020} & 21.6 &  &  & 2.9 &  & & 0.9 & 27.7 & 63.6 \\ 
CONUS \cite{obelix} & 50 & & 5 & & 7 & & 4 & 60 & 66 \\
\bottomrule
\end{tabular} \label{genat}
\end{table}

From all the results summarized in Tables~\ref{geenr} and \ref{genat} for germanium, it can be concluded that the order of the measured rates is in general compatible with calculations. For some isotopes, there is a considerable dispersion between different calculations following very different approaches and even between different experimental estimates, which could be related to differences in the flux and spectra of cosmic rays. In the case of tritium, the production rate is now very well estimated for natural germanium, as there is an excellent agreement between the measurements from EDELWEISS and CDMSlite and even with precise calculations \cite{tritiumpaper}. Considering specifically activation for DBD experiments, a good point is that cosmogenic activation is significantly suppressed for enriched germanium (with respect to natural germanium) for most of the isotopes, except for $^3$H (not affecting DBD experiments). The experimental determination of the production of $^{60}$Co in the detector medium is hard due to the simultaneous detection of beta and gamma emissions. Estimates of $^{68}$Ge are still subject to a great uncertainty due to the dispersion of results.

In addition to the germanium activation on the Earth's surface mainly due to cosmic nucleons, the underground in-situ activation has been considered for some DBD germanium projects. For the GERDA experiment, production rates of induced nuclides in detectors and materials of the set-up like the cyogenic liquid were firstly estimated from a MC simulation of muons in Gran Sasso using GEANT4 \cite{gerdamu}; the delayed decays of $^{77}$Ge and its metastable state $^{77m}$Ge following neutron captures in $^{76}$Ge were identified as the most relevant background. A re-evaluation of this cosmic muon induced background has been carried out for GERDA Phase II, obtaining a production rate of $^{77}$Ge/$^{77m}$Ge of $(0.21\pm0.01)$ nuclei per kg and year \cite{gerdamu2}. The estimated background contribution from this is well below the background level of GERDA Phase II and even below the one estimated for LEGEND thanks to the use of active background suppression techniques and the applications of delayed coincidence cuts. Other nuclides from cosmic muon interactions have been analyzed too finding a very minor contribution to the background.

Together with the quantification of the production of cosmogenic radioisotopes, the final effect of this induced activity in germanium experiments has been assessed too in general \cite{wei} and for particular projects. In Ref.~\cite{jian}, the expected spectra from the different cosmogenic radionuclides (taking into account phenomena produced by the coincidence summing-up effect) were evaluated from simulation for HPGe detectors used in rare event experiments.
\begin{itemize}
\item The effect of cosmogenic isotopes in germanium was computed in the background model developed for MAJORANA detectors \cite{majoranamodel}. It is shown that $^{68}$Ge gives always multiple-site events, which can be efficiently rejected through Pulse Shape Analysis (PSA). In addition, the low threshold of their detectors allows a further rejection applying time-correlation cuts with the $^{68}$Ge K,L-shell X-rays.
\item Following detailed analysis of the background \cite{gerdamodel,gerdamodel2}, the effect of $^{60}$Co and $^{68}$Ge was evaluated for GERDA concluding that they can be neglected when modeling Gerda Phase II data. The BEGe detectors were moved underground whenever possible during the fabrication and characterization and periods above ground were tracked. The expected impurities for Phase II data would produce 0.03 counts per day from $^{68}$Ge and 0.1~counts per day due to $^{60}$Co. The contribution for the detectors coming from the HM and IGEX experiments is estimated to be even smaller thanks to the long storage underground. According to simulations, the background contributions in the region of the transition energy are less than 10$^{-4}$~counts keV$^{-1}$ kg$^{-1}$ y$^{-1}$ in both cases.
\item Based on the validated codes against CDEX-1B detector data, a prediction was made on the background level from cosmogenics for the tonne-scale CDEX experiment \cite{cdexcos}, being of the order of 10$^{-7}$~counts keV$^{-1}$ kg$^{-1}$ d$^{-1}$ around 2 MeV assuming reasonable times for exposure on surface and cooling at the Jinping underground laboratory. For dark matter searches, it is considered that crystal growth and detector manufacture should inevitably be made underground due to tritium and X-ray emissions. In germanium detectors having efficient discrimination between nuclear and electronic recoils, like SuperCDMS, it helps to reduce the effect of cosmogenics \cite{wei}.
\end{itemize}

\section{Tellurium}
\label{secte}

Tellurium has also been considered in different DBD experiments for years and therefore cosmogenic activation in tellurium compounds has been carefully analyzed too. As for germanium, in this section a brief overview of tellurium DBD experiments is given before the description of activation studies.

\subsection{Tellurium DBD experiments}

 The DBD of $^{130}$Te has been largely studied using bolometers \cite{cremonesi18,brofferio18,brofferio19}. Thanks to a high natural isotopic abundance of 34\%, enrichment is not necessary which makes it feasible to accumulate large amounts of tellurium oxide. Tellurium compounds show a good radiopurity but working as cryogenic detectors, operation at very low temperatures (even below a few tens of mK) is required. The transition energy of this isotope (around 2527~keV \cite{qte1,qte2,qte3}) is quite high. Tellurite bolometers have shown excellent energy resolution, at the level of 5 to 10~keV for FWHM at the transition energy. Surface contamination is a problematic background for this type of detectors and different approaches have been followed to suppress it; scintillating bolometers (measuring simultaneously heat and light) have been developed with particle identification capabilities, which allows to efficiently reject surface events. In addition, alpha discrimination is also possible in tellurite detectors from the simultaneous measurement of the Cherenkov light generated by gamma/beta emissions \cite{berge,tabarelli}.

Experiments with increased mass of TeO$_{2}$ and sensitivity have been successively installed at the LNGS for DBD searches, from MIBETA in the nineties \cite{ALE00c} to the present ``Cryogenic Underground Observatory for Rare Events'' (CUORE) with 741~kg \cite{cuore18}, including CUORICINO with 40.7~kg \cite{prccuoricino} and CUORE-0 with 39~kg  \cite{prlcuore0}. CUORE is operating 988 cubic crystals (side 5 cm each, grouped in 19 towers) with a total mass of $^{130}$Te of 206~kg. The operation temperature is $\sim$10~mK, which is achieved using a cryogen-free dilution refrigerator. A FWHM energy resolution of $(7.7\pm0.5)$~keV has been measured and in the region of interest the background level is $(0.014\pm0.002)$ counts keV$^{-1}$ kg$^{-1}$ y$^{-1}$ \cite{cuore18}, one order of magnitude lower than in CUORICINO. The latest results at 90\% C.L. corresponding to an exposure of of 372.5 kg$\cdot$y are $T_{1/2}^{0\nu}\geq 3.2 \times 10^{25}$~y for the neutrinoless DBD half-life of $^{130}$Te and  $m_{\beta\beta} \leq 75-350$~meV, depending on the nuclear matrix elements, for the effective neutrino mass \cite{cuore19}. Limits for other DBD modes of $^{130}$Te have been also derived from CUORE-0 \cite{cuore0exc,cuore0EC} .

CROSS (``Cryogenic Rare-event Observatory with Surface Sensitivity'') foresees the installation of  arrays of both enriched Li$_{2}$MoO$_{4}$ and TeO$_{2}$ bolometers to investigate the DBD of $^{100}$Mo and $^{130}$Te \cite{cross}. It is in commissioning phase at the Canfranc Underground Laboratory \cite{lscppnp}. Pulse shape discrimination techniques based on Solid-State-Physics phenomena are in development to identify and reject not only alpha but also beta surface contamination. Unlike scintillating bolometers, CROSS detectors operate without light detectors, which simplifies the bolometric set-up.

$^{130}$Te has been also investigated using other detection technologies. Large neutrino detectors used for oscillation studies based liquid scintillators can also be considered to investigate DBD thanks to the techniques developed for loading nuclei into the liquid scintillator. This type of detectors have poor energy resolution (increasing the leakage of the two-neutrino DBD signal into the neutrinoless peak) but event reconstruction is possible and fiducial volumes can be defined to reduce background \cite{reviewchen}. Once the Sudbury Neutrino Observatory (SNO) experiment for  solar neutrino measurements at SNOLAB was finished, the heavy water used was replaced by a liquid scintillator (linear alkylbenzene with PPO) to start the multi-purpose SNO+ experiment. $^{150}$Nd was initially considered as dopant for DBD searches, but finally $^{130}$Te was chosen and the plan is to load 1.3~tonnes of the isotope from natural tellurium. Some data have been taken with ultrapure water \cite{sno}.

\subsection{Activation studies}

In this context, activation studies for tellurium has been made mainly focused on CUORE and SNO+. Irradiation measurements using proton and neutron beams and different calculations have been made. As highlighted when producing the background model for CUORE \cite{cuorebkg}, the cosmogenic isotopes relevant to the searches of the DBD of $^{130}$Te must have a greater transition energy, relevant  production cross sections and half-lives longer than the time scale of the experiment. $^{110m}$Ag and $^{60}$Co fulfill these criteria. To affect the region of the neutrinoless DBD signal, both the gamma and beta emissions from $^{60}$Co must leave their energy in a crystal. $^{110m}$Ag is a $\beta^-$ emitter with transition energy of 3009.8 keV and half-life 249.8 days. Although not relevant for DBD, the tritium production in TeO$_{2}$ has been quantified using TALYS \cite{mei}.

Cross sections for proton production were measured within CUORE both in US and Europe \cite{te}, after first results on proton spallation on tellurium \cite{bardayan,norman}. A tellurium target was irradiated at LANSCE by a 800~MeV proton beam and later gamma screened with germanium detectors at Berkeley. Targets made of TeO$_{2}$ were also exposed to proton beams (of 1.4 and 23~GeV) at CERN; the gamma analysis with germanium detectors was carried out first at CERN and then in Milano several years later. The obtained results for the three proton energies agree reasonably with the estimates from the Silberberg\&Tsao formulas.
In addition, TeO$_{2}$ powder was irradiated also at LANSCE but with the neutron beam with energies up to $\sim$800 MeV resembling the cosmic neutron spectrum at sea level \cite{ten}. After gamma analysis at Berkeley, production cross sections were derived. Production rates for $^{110m}$Ag and $^{60}$Co were deduced, assuming the Gordon et al. neutron spectrum, from the measured cross sections, as reported in Table~\ref{te}.

Activation yields for $^{130}$Te experiments under certain exposure conditions have been also computed using COSMO in the context of other bolometers in Ref. \cite{cosmobol}, as it will be later described in Sec. \ref{secotherDBD}.

In the context of SNO+, production rates of many radioisotopes induced on natural tellurium were evaluated using ACTIVIA (above 100~MeV) and the TENDL library (in 10-200~MeV when available) and considering the parameterization for cosmic nucleons from \cite{armstrong,gehrels}, as presented in Ref.~\cite{telozza}; some of these results are presented in Table~\ref{te}. A significant discrepancy is observed between the different estimates of the rates for $^{110m}$Ag and $^{60}$Co from \cite{ten} and \cite{telozza}, although the studies are not directly comparable and cross sections above 800~MeV are in good agreement.

\begin{table}
\caption{Rates of production at sea level of long-lived radioisotopes deduced from measured cross sections for TeO$_{2}$ in Ref.~\cite{ten} and from calculations for natural tellurium in Ref.~\cite{telozza} (see text). All numbers are expressed in atoms/kg/day.}
\centering
\begin{tabular}{@{}lccc@{}}
\toprule
  & \textbf{$^{60}$Co} & \textbf{$^{110m}$Ag} & \textbf{$^{124}$Sb} \\
\midrule
Half-life \cite{ddep}  & 5.2711(8)~y & 249.78(2)~d & 60.208(11)~d \\ \hline
Measurement \cite{ten} & $<$0.0053 & 0.42 &  \\
ACTIVIA+TENDL \cite{telozza} & 0.070 & 0.206 & 15.7 \\
\bottomrule 
\end{tabular} \label{te}
\end{table}

The underground activation for natural tellurium was also studied in Ref.~\cite{telozza}. Muon-induced and radiogenic neutrons from ($\alpha$,n) reactions were considered. For the depth of SNOLAB ($\sim$6 km.w.e.) and considering long-lived isotopes, estimates pointed to $<$1 event/(y t). For short-lived radioisotopes of Sn, Sb and Te, the production rates derived from ACTIVIA and TENLD data are orders of magnitude lower than the corresponding rates on surface.

The effect of cosmogenic activation in the background of the CUORE experiment has been carefully quantified. To minimize the activation levels in the TeO$_2$ crystals of CUORE, exposure to cosmic rays was carefully controlled: crystals were stored underground at LNGS just three months after the growth and spent on average four years for cooling before use. From the neutron and proton irradiation measurements described \cite{te,ten}, the activation levels in the crystals (one year before starting the data taking of CUORE) were evaluated as $<$20 nBq/kg of $^{110m}$Ag/$^{110}$Ag and $<$1 nBq/kg of $^{60}$Co. In the region of interest, the corresponding estimated background level from the cosmogenic activation of TeO$_2$ was $<$6.7$\times 10^{-5}$ counts keV$^{-1}$ kg$^{-1}$ y$^{-1}$, much lower than the total projected background around $10^{-2}$ counts keV$^{-1}$ kg$^{-1}$ y$^{-1}$ \cite{cuorebkg}. Indeed, in the detailed reconstruction of the sources of the CUORE-0 counting rate to obtain the two-neutrino DBD  half-life of $^{130}$Te \cite{cuore02nu} the only cosmogenic isotope in the crystals quantified through its gamma emissions was $^{125}$Sb, not relevant for the neutrinoless DBD as it is a $\beta^-$ emitter with a transition energy of 766.7 keV.

\section{Xenon}
\label{secxe}

Xenon is also a very relevant target in present and future DBD experiments. Activation studies for this material have been made; as it will be shown in this section, although spallation products are not an issue in general in these experiments, the effect of induced $^{137}$Xe is being carefully considered.

\subsection{Xenon DBD experiments}

The detectors of experiments described in Secs.~\ref{secge} and \ref{secte} register the energy of the two electrons emitted in the DBD; but using different technologies, the electron tracks can be measured too. A signal event from DBD should show two short tracks with a common origin while background events have a different topology: muons give longer tracks and beta/gamma emission generate energy deposits at separate positions. In the approach followed by the SuperNEMO project
\cite{supernemoepjc}, the DBD source is a thin sheet external to the detector, consisting of gas detectors for tracking and scintillators as calorimeters; it follows the successful  ``Neutrino Ettore Majorana Observatory'' (NEMO) experiment at  the Modane Underground Laboratory, which has released results for several DBD emitters \cite{nemoca,nemose,nemomo,nemond}. Other approach to develop tracking detectors for DBD is based on TPCs, in particular filled with xenon (either gas or liquid) to investigate $^{136}$Xe \cite{reviewmichel,revHPXe}. Tracks are obtained from the ionization while the energy is deduced from scintillation. $^{136}$Xe is a very attractive DBD emitter for several reasons: enrichment (natural isotopic abundance of $^{136}$Xe is 8.86\%) is feasible; the transition energy ($(2457.83\pm0.37)$~keV \cite{QXe}) is quite high; xenon is radiopure and can be purified; and the two-neutrino mode is slow. The energy resolution is better in gas than in liquid TPCs, but both techniques are being implemented.

``Enriched Xenon Observatory'' (EXO), located at WIPP, in New Mexico, US, has developed liquid xenon TPCs. The detector of EXO-200 contained 200 kg of enriched xenon (80.6\% of $^{136}$Xe) having wire planes and avalanche photodiodes at the two ends of the cylindrical chamber to read charge and light, respectively; it was built from carefully selected radiopure materials \cite{exoscreening} and produced important results \cite{exonature}. For the two-neutrino DBD channel of $^{136}$Xe, a half-life shorter than expectations was confirmed \cite{exo2nu} while for the neutrinoless mode the 90\% C.L. limit derived is $T_{1/2}^{0\nu}>3.5 \times 10^{25}$~y \cite{exoprl}. Other modes and searches have also been explored  \cite{exomaj,exoexc,exo134,exonuc}. The nEXO observatory (with 5~tonnes of enriched xenon) is the next phase of the project, with the aim to improve the sensitivity two orders of magnitude \cite{nexo}.
Barium tagging, intended to individually identify the Ba atoms produced by the DBD, is being studied \cite{exotag}.

The technology of high-pressure xenon gas (HPXe) TPCs with electroluminescent amplification is being developed to investigate the neutrinoless DBD of $^{136}$Xe by NEXT (``Neutrino Experiment with a Xenon TPC'') in the Canfranc Underground Laboratory \cite{lscppnp}. Several prototypes (with mass of $\sim$1~kg) were designed and operated in a first step to demonstrate the advantages of the technology, including a distinctive signal topology and very good energy resolution \cite{berkeley,valencia}. As a second phase, the NEXT-White demonstrator (with 5~kg of xenon) was built in Canfranc \cite{nextnew} and in 2020 is running smoothly, having confirmed the background discrimination capability from the topological signature \cite{recognitionII} and shown an energy resolution of 1\% (FWHM) in the region of interest \cite{resolutionII}; energy resolution, which depends on the stability of operation parameters, wave-shifters, light detectors and other elements, is much better in xenon gas than in liquid because the fluctuations in the ionization production are smaller than the ones due to pure Poisson statistics (Fano factor is lower than 1 in gaseous phase). NEXT-100 (with 100~kg of xenon at 15~bar) is the next stage of the program \cite{nextsensitivity}, built with radiopure specifications \cite{nextradiogenicbkg} as a scale up of NEXT-White by 2:1 in size; operation might start in 2021. In the longer term, for a tonne-scale detector exploring neutrinoless DBD half-lives beyond 10$^{27}$~y, NEXT-HD follows the same technology while NEXT-BOLD proposes the implementation of the detection of Ba ions which is under investigation \cite{batagging,bicolor}. NEXT, as CUPID and LEGEND, is recognized by the Double Beta Decay APPEC Committee as a competitive project \cite{appec}.

Based also on HPXe TPCs, other projects pursue the search for the neutrinoless DBD of $^{136}$Xe. The PandaX-III (``Particle And Astrophysical Xenon Experiment III'') experiment is working to build firstly a TPC with Microbulk Micromegas readout for tracking and 200~kg of enriched xenon (90\% in $^{136}$Xe) operated at 10~bar in the Jinping laboratory \cite{pandaxiii}. The background discrimination from track analysis has been studied \cite{pandaxtopo}. As a second step, the implementation of five of such TPCs is foreseen. In Japan,  the AXEL (A Xenon ElectroLuminescence) project explores a new system for the collection of the electroluminescence light \cite{axel} and prototypes are in development.

As pointed out in Sec.~\ref{secte} for $^{130}$Te, large liquid-scintillator detectors loaded with DBD emitters have joined the investigation of DBD. A nylon balloon filled with 13~tons of a Xe-loaded liquid scintillator and read by photomultiplier tubes (PMTs) was inserted inside the KamLAND detector, at the Kamioka mine in Japan, to carry out the KamLAND-ZEN experiment. About 320~kg of enriched xenon gas (90.9\% of $^{136}$Xe) were loaded. After the reduction of the $^{110m}$Ag contamination firstly identified, the 90\% C.L. limits derived were $T_{1/2}^{0\nu}>1.07 \times 10^{26}$~y for the neutrinoless DBD  half-life of $^{136}$Xe and 61-165~meV (according to common nuclear matrix element) for the effective neutrino mass \cite{kamlandzen0nu}. Results for the two-neutrino channel \cite{kamlandzen2nu} and other modes \cite{kamlandzenexc,kamlandzenmaj} have been also derived. A new phase of the experiment with 745~kg of xenon in a larger balloon is underway.

\subsection{Activation studies}

Xenon-based detectors play also an essential role for the direct detection of dark matter, like in XENON1T \cite{xenon1t}, LUX-ZEPLIN \cite{lztdr,lz}, PANDAX \cite{pandax} and XMASS \cite{xmass} and future DARWIN \cite{darwin}. Sensitivity of some of these projects also to the neutrinoless DBD of $^{136}$Xe has been evaluated, taking advantage of the significant amount of this isotope even without isotopic enrichment \cite{darwindbd,lzdbd}. Purification systems for the liquid xenon are supposed to remove all non-noble radionuclides, but studies of cosmogenic activation in xenon have been made from different calculations and measurements. Indeed, tritium has been considered as a possible explanation for the excess of electronic recoil events observed in XENON1T below 7 keV \cite{xenon1texcess,robinson}. In-situ calibrations performed in some experiments using neutrons with energies of some MeV are not problematic due to the different energy ranges of neutrons producing relevant activation; $^{129m}$Xe and $^{131m}$Xe are induced, but they have short half-lives of 8.88 and 11.96 days, respectively. Spallation products identified in xenon studies are not relevant for neutrinoless DBD searches but $^{137}$Xe, produced by neutron capture on $^{136}$Xe, has been found as a significant background in DBD experiments. This isotope is a $\beta^-$ emitter having a transition energy of 4173 keV and a half-life of 3.82 minutes; it creates coincident gamma emissions only 33\% of the decays.

A dedicated measurement of cosmogenic yields in xenon (and copper) is described in Ref.~\cite{baudis}. The controlled exposure of the samples to cosmic rays was carried out for 345~days at the Jungfraujoch research station in Switzerland (at an altitude of 3470~m) and gamma screening was made, before and after activation, with a sensitive germanium detector at LNGS. ACTIVIA and COSMO, with cosmic neutron spectrum from \cite{armstrong,gehrels}, were used to perform calculations of the measured production rates. Activity of 18 radionuclides was quantified or constrained; some of the derived results are presented in Table~\ref{xe}. Among the identified radioisotopes, only $^{125}$Sb was finally considered in this work as a possible background for xenon dark matter searches.

Estimates of production rates of radioisotopes induced in natural xenon were made, as for germanium, using TALYS in Ref.~\cite{mei} and, as for other materials, using GEANT4 (with the Shielding modular physics list) and ACTIVIA (considering semiempirical formulae and data tables) in Ref.~\cite{mei2016}. Calculations presented in Table~\ref{xe}, including $^{3}$H, correspond to those made with the Gordon et al. cosmic neutron spectrum. In Ref.~\cite{mei2016}, a production rate of $^{127}$Xe of about 3~atoms per tonne and per day was calculated using GEANT4 and taking into account thermal and fast neutrons at SURF (at a depth of 4.3 km.w.e., 1480~m); it is considered to be irrelevant.

The cosmogenic yields of some nuclides were quantified from the background analysis of the data of the LUX experiment, operated at SURF in the US, taken at different times \cite{lux}. Table~\ref{xe} shows some of the results, as quoted in Ref.~\cite{baudis}. For the LUX-ZEPLIN project, it is concluded that, once xenon is underground, the cosmogenic activity becomes negligible.

\begin{table}
\caption{Rates of production of long-lived radioisotopes induced in natural xenon at sea level. Measurements from Refs.~\cite{baudis,lux} and different calculations are presented (see text). All numbers are expressed in atoms/kg/day.}
\centering
\begin{tabular}{@{}lcccccc@{}}
\toprule
& \textbf{$^{3}$H} & \textbf{$^{7}$Be} & \textbf{$^{125}$Sb} & \textbf{$^{121m}$Te} & \textbf{$^{123m}$Te} & \textbf{$^{127}$Xe} \\
\midrule
Half-life \cite{toi,ddep} & 12.312(25)~y & 53.22(6)~d & 2.75855(25)~y & 154~d & 119.3(1)~d & 36.358(31)~d  \\ \hline
Measurement \cite{baudis} & & 32$^{+21}_{-20}$ &  51$^{+22}_{-20}$ & $<$104 & $<$53 & 162$^{+25}_{-23}$  \\
Measurement \cite{lux} & & & & & & 132$\pm$26  \\
COSMO \cite{baudis} & & 0.55 & 1.17 & 23.8 & 1.24 & 48.0  \\
ACTIVIA \cite{baudis} & & 0.55 & 0.017 & 25.8 & 1.27 & 35.9 \\
ACTIVIA \cite{mei2016} & 35.6 & & 0.009 & 54.5 & 2.67 & 89.9 \\
GEANT4 \cite{mei2016} & 31.6 & & 1.48 & 21.2 & 18.5 & 233.3 \\
TALYS \cite{mei} & 16.0 &  & 0.04 & 11.7 & 12.1 & \\
\bottomrule
\end{tabular} \label{xe}
\end{table}

A thorough study of the background due to muon activation underground was made for the EXO-200 experiment \cite{exobkg} based on both experimental data and simulations using FLUKA and GEANT4. From the simulations, cosmogenic nuclide production rates were derived taking into account the measured muon flux at WIPP as determined by the EXO-200 TPC, identifying potential worrisome products. The study of veto-tagged data (searching for coincidences between signals of neutron capture in TPC and muon veto triggers) allowed to obtain the rates of neutron capture in the detector. Considering radioisotopes which can contribute to the region of interest for neutrinoless DBD, the production of several iodine isotopes and of $^{135}$Xe and $^{137}$Xe was computed; iodine atoms must be eliminated by the xenon purification and $^{135}$Xe cannot mimic the neutrinoless DBD signal. Therefore, only $^{137}$Xe was found to have a significant contribution. Its production rate in EXO-200 was estimated as $439\pm17$ ($403\pm16$) atoms per year from GEANT4 (FLUKA) simulation. The measured capture rate deduced is 338$^{+132}_{-93}$ captures on $^{136}$Xe per year, in agreement within uncertainties with the expectation. For EXO-200 (considering data from 123.7 kg$\cdot$y), in the neutrinoless DBD region of interest, $^{137}$Xe is considered to contribute with 7.0 out of a total 31.1 counts \cite{exobkg}. This very relevant contribution can force future liquid xenon TPCs to go deeper underground.

Production of $^{137}$Xe is being deeply considered also in other experiments. In KamLAND-ZEN, based on the spallation neutron rate and the $^{136}$Xe capture cross section, the production yield is estimated to be ($3.9\pm2.0)\times10^{-3}$ (tonne$\cdot$day)$^{-1}$, consistent with their simulation study using FLUKA \cite{kamlandzen0nu}. Contribution to background in the NEXT experiment is being also computed; moreover, a method has been proposed to mitigate cosmogenic $^{137}$Xe using $^3$He \cite{mitigation}. This study, based on GEANT4 simulations, considers the addition of small quantity of $^3$He to xenon in order to capture thermal neutrons and then decrease the detector activation. Activation rates and the corresponding background level in the region of interest have been computed for several labs placed at different depths (LSC, LNGS, SURF and SNOLAB) for different percentages of $^3$He. For instance, for LNGS, an activation rate of 0.10 kg$^{-1}$ y$^{-1}$ is estimated for operation of pure enriched xenon, with a reduction of a factor 10 with the addition of 0.1\% $^3$He by mass. It is concluded that, thanks to $^{136}$Xe/$^3$He mixtures, the effect of $^{137}$Xe activation on neutrinoless DBD searches can be made negligible for HPXe detectors at the tonne scale and beyond operating at any underground laboratory \cite{mitigation}.

In addition, it is worth noting that an inverse kinematics experiment carried ut at GSI \cite{xe136} deduced with high accuracy for $^{136}$Xe and for 271 medium-mass radioisotopes the corresponding production cross sections. Nuclei were created by spallation when launching $^{136}$Xe projectiles on a liquid hydrogen target (at 500 MeV per nucleon) and were unambiguously identified using a magnetic spectrometer. Cross sections for Ba, Cs, I, Sb, Sn, Te and Xe have been derived.

\section{Other DBD target materials}
\label{secotherDBD}

Together with $^{76}$Ge, $^{130}$Te and $^{136}$Xe, other DBD emitters are being investigated in different projects. As mentioned in Sec. \ref{secxe}, the NEMO3 experiment has studied $^{82}$Se, $^{100}$Mo and $^{150}$Nd among others and much larger masses of $^{82}$Se and $^{150}$Nd are planned to be used in SuperNEMO. Profiting from the experience at LUCIFER \cite{lucifer}, CUPID-0 has become the first experiment operating a large array of scintillating bolometers made of ZnSe with particle identification capabilities. The CUPID-0 detector has five towers with 26 ZnSe scintillating crystals and a total detector mass of 10.5 kg operating at $\sim$10 mK at LNGS. The crystals are interleaved with germanium light detectors, being all equipped with NTD Ge thermistors. The 90\% C.L. limit for the neutrinoless DBD half-life of $^{82}$Se is $T_{1/2}^{0\nu}\geq 3.5 \times 10^{24}$~y, corresponding to an effective neutrino mass $m_{\beta\beta} \leq 311-638$~meV, following different nuclear matrix element calculations \cite{cupid0}. Regarding the DBD studies of $^{100}$Mo, crystals made of ZnMoO$_{4}$, Li$_{2}$MoO$_{4}$ and CaMoO$_{4}$ are used by LUMINEU and CUPID-Mo \cite{lumineu,cupidmo}, CLYMENE \cite{clymene}, CROSS \cite{cross} and AMoRE \cite{amore} experiments. $^{150}$Nd was considered in the SNO+ experiment, as pointed out in Sec. \ref{secte}, although it was finally disregarded. Two-neutrino DBD to excited states has been observed for both $^{100}$Mo \cite{excmo} and $^{150}$Nd \cite{excnd}. Other DBD emitters like $^{116}$Cd can be studied using low background scintillators \cite{barabash2020}. Even if limited in some cases, there is also information on the cosmogenic activation of targets containing some of these DBD isotopes.

Production cross sections for natural neodymium induced by protons with energies from 10 to 30~MeV were obtained in Ref.~\cite{nd1}, following an irradiation made at a cyclotron in Rez, in Czech Republic, and a later gamma analysis with a germanium detector. Comparisons with calculations using TENDL-2010 give a reasonable agreement in values and trends. Considering a proton flux \cite{barabanov}, production rates of radioisotopes of relevance for DBD were computed. The study was completed in Ref.~\cite{nd2}, considering energies from 5 to 35~MeV and TENDL-2012 results. Table \ref{nd} summarizes the measured and calculated production rates of several long-lived isotopes, decaying mostly by electron capture and giving gamma emission not relevant for neutrinoless DBD.

\begin{table}
\caption{Rates of production of long-lived radioisotopes induced in neodymium by protons at sea level. Measurements from Refs.~\cite{nd1,nd2} and calculations are presented (see text). All numbers are expressed in atoms/kg/day.}
\centering
\begin{tabular}{@{}lccc@{}}
\toprule
& \textbf{$^{143}$Pm} & \textbf{$^{144}$Pm} & \textbf{$^{146}$Pm}  \\
\midrule
Half-life \cite{toi} & 265~d & 363~d & 5.53~y   \\ \hline
Measurement \cite{nd1} & 0.1260 & 0.0830 & 0.0196 \\
TENDL \cite{nd1} & 0.1665 & 0.0967 & 0.0255 \\
Measurement \cite{nd2} & 0.1753 & 0.1092 & 0.0276 \\
TENDL \cite{nd1} & 0.2187 & 0.1196 & 0.356 \\
\bottomrule
\end{tabular} \label{nd}
\end{table}

A very detailed study of cosmogenic activation for materials used as bolometers in DBD experiments was presented in Ref. \cite{cosmobol}. $^{65}$Zn is typically observed in detectors containing Zn, $^{75}$Se was registered in ZnSe and $^{116}$CdWO$_4$ exhibits $^{110m}$Ag. Activation by $^{22}$Na is reported for calcium. Using COSMO and considering as target elements of selected  scintillators as well as copper, initial expected decay rates, considering one month of sea level exposure and another month of cooling underground, have been computed for relevant isotopes with half-life longer than 30 days; in these conditions, the highest (initial) activity found from cosmogenics is at the level of a few to tens of $\mu$Bq/kg. As the quantified products decay in many cases to excited states, the generated events can be identified looking for coincidences and efficiently suppressed. In any case, it is recommended for all the analyzed targets strictly controlling this activation.

The cosmogenic activation of Li$_2$MoO$_4$ crystals was specifically estimated for CUPID using the ACTIVIA code, assuming certain exposure conditions \cite{cupidprecdr}. The only potentially dangerous activated isotopes found in $^{100}$Mo are $^{82}$Rb, $^{56}$Co and $^{88}$Y. The contribution of these  isotopes to the CUPID background estimated by MC simulation is lower than  5$\times10^{-5}$ counts/(kg·day). In addition, the induced activity of cosmogenic $^{65}$Zn in ZnSe detectors has been quantified when developing the background model of the CUPID-0 experiment \cite{cupid0bkg}, while no other cosmogenic isotope has been identified for these crystals.

Finally, it is worth noting that the same irradiation made for germanium and tellurium at LANSCE with protons of 800~MeV was also performed on a natural molybdenum sample; the production cross section of $^{60}$Co was determined after gamma counting at Berkeley \cite{norman}.

\section{Other non-DBD target materials}
\label{othernonDBD}

Cosmogenic activation for DBD experiments is relevant not only for target materials in the detectors but also for materials used in components, shielding and ancillary systems in the whole set-up. Activation studies have been made in the context of rare event experiments for example for copper, lead, stainless steel, titanium, aluminum and argon; they will be summarized in this section. Some of the identified cosmogenic products in these materials are $\beta$ emitters with high transition energies producing gamma emissions which can reach the detector and affect the region of interest for some neutrinoless DBD searches.

\subsection{Copper}
\label{seccu}

Copper is widely used in rare event experiments in components of different types of detectors and in shieldings, thanks to its advantageous electrical, thermal and mechanical properties. In addition, for this material there is a large amount of production cross sections measured with proton and even neutron beams, which makes copper an optimal material for validation of calculations. Consequently, cosmogenic activation of copper has been quite extensively considered.

Several studies are based on calculations from codes following MC simulations or semiempirical equations. Production rates from these studies are shown in Table~\ref{cu}.
\begin{itemize}
\item ACTIVIA was applied to calculate for benchmark production rates in copper, as also made for germanium \cite{activia} (see Sec. \ref{secge}).
\item As described also for other targets, different production rates were computed using TALYS in Ref.~\cite{mei} and GEANT4 and ACTIVIA in Ref.~\cite{mei2016}.
\item The study in Ref.~\cite{cebrian} (obtaining production rates from selected excitation functions according to deviation factors) applied to germanium (see Sec. \ref{secge}) was also made for copper. For this material, rates were calculated below/above 100~MeV using, respectively, the MENDL2N library and YIELDX calculations combined with experimental data. Results shown in Table~\ref{cu} were obtained considering the Gordon et al. spectrum.
\item The calculations using GEANT4 and CONUS software packages and ACTIVIA to obtain production rates of radioisotopes and estimate the corresponding counting rates in the Obelix HPGe detector \cite{obelix} (see Sec. \ref{secge}) included also copper as it is used in the cold finger.
\end{itemize}

In addition to computed production rates, for copper there are also direct estimates carried out from natural irradiation of material.
\begin{itemize}
\item A first direct measurement of saturation activities was presented in Ref.~\cite{lngsexposure}. The exposure to cosmic rays of the copper samples (total mass 125~kg) provided by Norddeutsche Affinerie (now Aurubis, in Germany) took place for 270~days at Gran Sasso (altitude 985~m). A long (103~days) gamma screening was made at LNGS using the GeMPI detector. Cobalt isotopes gave the highest yields, finding for $^{60}$Co an activity much higher than the initial one. The production rates from this work presented in Table~\ref{cu} have been derived for sea level by taking into account an altitude correction factor of 2.1 (estimated as in Ref.~\cite{ziegler}).
\item Together with xenon (see Sec. \ref{secxe}), copper was analyzed in the same way too in the study of Ref.~\cite{baudis}. Several OFHC copper samples (provided by Norddeutsche Affinerie, from a batch used in the construction of some components of XENON100) with a total mass of 10.35~kg were exposed at the same place that the xenon sample also for 345~day. Germanium gamma analysis at LNGS, before and after activation, was made using the Gator detector to quantify saturation activities. Table~\ref{cu} shows the results from the measurements as well as those from ACTIVIA and COSMO estimates. For the XENON1T detector, a complete material radioassay was made; activities of $^{54}$Mn and $^{55-58}$Co were quantified from HPGe spectrometers in different copper samples, finding variation from batch to batch, depending on the storage and shipment of the material \cite{xenon1trpurity}.
\item Activation in copper used for shielding was studied in Ref.~\cite{ivan}. A sample (with mass of 18~kg) was exposed to cosmic rays for 1~year at an altitude of 250~m and copper bricks for 41~days. Following the germanium measurements performed at LSC, the derived activities for $^{54}$Mn and different cobalt radioisotopes are in good agreement with predictions from production rates.
\end{itemize}

\begin{table}
\caption{Rates of production of long-lived radioisotopes induced in natural copper at sea level. Measurements from Refs.~\cite{lngsexposure,baudis} and different calculations are presented (see text). Several cosmic neutron spectra have been assumed in the ACTIVIA calculations from different works. All numbers are expressed in atoms/kg/day.}
\centering
\begin{tabular}{@{}lcccccccc@{}}
\toprule &
\textbf{$^{46}$Sc} & \textbf{$^{48}$V} & \textbf{$^{54}$Mn} &  \textbf{$^{56}$Co} & \textbf{$^{57}$Co} & \textbf{$^{58}$Co} & \textbf{$^{59}$Fe} & \textbf{$^{60}$Co} \\
\midrule
Half-life\cite{toi,ddep} & 83.787(16) & 15.9735 & 312.19(3) & 77.236 & 271.81(4) & 70.85(3) & 44.494 &  5.2711(8)   \\
units & d & d & d & d & d & d & d & y \\ \hline
Measurement \cite{lngsexposure} & 2.18$\pm$0.74 & 4.5$\pm$1.6 & 8.85$\pm$0.86 &   9.5$\pm$1.2 & 74$\pm$17 & 67.9$\pm$3.7 & 18.7$\pm$4.9 & 86.4$\pm$7.8 \\
Measurement \cite{baudis} & 2.33$^{+0.95}_{-0.78}$ & 3.4$^{+1.6}_{-1.3}$ & 13.3$^{+3.0}_{-2.9}$ &  9.3$^{+1.2}_{-1.4}$ & 44.8$^{+8.6}_{-8.2}$ & 68.9$^{+5.4}_{-5.0}$ & 4.1$^{+1.4}_{-1.2}$ & 29.4$^{+7.1}_{-5.9}$ \\
ACTIVIA (MENDL-2P) \cite{activia} & 3.1 & & 12.4 & 14.1 & 36.4 & 38.1 & 1.8 &  9.7 \\
ACTIVIA \cite{activia,baudis} & 3.1 & & 14.3 &  8.7 & 32.5 & 56.6 & 4.2 & 26.3 \\
COSMO \cite{baudis} & 1.5 & 3.1 & 13.5 &  7.0 & 30.2 & 54.6 & 4.3 &  25.7 \\
ACTIVIA \cite{mei2016} & 4.1 & & 30.0 &   20.1 & 77.5 & 138.1 & 10.5 & 66.1 \\
ACTIVIA \cite{obelix} & 3 & & 16 &  9 & 34 & 60 & 2 & 29 \\
GEANT4 \cite{mei2016} & 1.2 & & 12.3 &  10.3 & 67.2 & 57.3 & 8.8 & 64.6 \\
TALYS \cite{mei} & & &  16.2 &   & 56.2 &  &  & 46.4 \\
MENDL+YIELDX \cite{cebrian} &  2.7 &  & 27.7 &  20.0 & 74.1 & 123.0 &   4.9 & 55.4 \\
CONUS \cite{obelix} & 3 & & 14 &  10 & 50 & 76 & 5 & 92 \\
 \bottomrule
\end{tabular} \label{cu}
\end{table}

As it can be deduced from Table~\ref{cu}, the two sets of measured production rates are compatible except for some nuclides like $^{60}$Co, which is probably the most relevant cosmogenic product in copper for DBD searches. The rates from Ref.~\cite{lngsexposure} are usually higher. Discrepancies between calculations are in some cases important.

As made for activation in tellurium (see Sec. \ref{secte}), cosmogenic activation in copper has been carefully considered in the CUORE experiment.
The final copper components which are placed close to the crystals were stored underground after four months of the production of the raw material. The activation level of $^{60}$Co for these components was evaluated (for one year before the starting of the data taking of CUORE) as $<$35 $\mu$Bq/kg, which agrees with the results of the germanium screening of the same material ($<$25 $\mu$Bq/kg) \cite{cuorebkg}. The counting rated estimated from this copper activation in the region of interest was $<$2.2$\times 10^{-6}$~counts keV$^{-1}$ kg$^{-1}$ y$^{-1}$, lower than the one predicted for tellurium activation \cite{cuorebkg}. In the reconstruction of background components for the CUORE-0 to obtain the two-neutrino DBD half-life of $^{130}$Te \cite{cuore02nu}, $^{54}$Mn, $^{57}$Co and $^{60}$Co produced in copper were quantified through their gamma emissions.

\subsection{Lead}
\label{secpb}

Although large amounts of lead are use for shielding in experiments looking for rare events, there are not many studies on its cosmogenic activation. As presented in Ref.~\cite{giuseppe}, the irradiation of a lead sample was performed at LANSCE with the neutron beam which mimics the cosmic neutron spectrum; germanium gamma counting was made at WIPP to finally derive the production rates of some isotopes at sea level. TALYS calculations using the Gordon et al. cosmic neutron spectrum were also made for comparison. Some of these results are presented in Table~\ref{pb}. From this work, it was concluded that assuming usual exposure conditions, the background generated by cosmogenics is less relevant than the one from common radionuclides present in lead.

\begin{table}
\caption{Rates of production of long-lived radioisotopes induced in natural lead at sea level. Measurements and calculations from Ref.~\cite{giuseppe} are presented (see text). All numbers are expressed in atoms/kg/day.}
\centering
\begin{tabular}{@{}lccc@{}}
\toprule
 & \textbf{$^{194}$Hg} & \textbf{$^{202}$Pb} & \textbf{$^{207}$Bi} \\
\midrule
Half-life (y) \cite{toi,ddep} & 444 &  5.25 10$^{4}$  & 32.9  \\ \hline
Measurement \cite{giuseppe} & 8.0$\pm$1.3 & 120$\pm$25 &  $<$0.17  \\
TALYS \cite{giuseppe} &  16 & 77 & \\
\bottomrule 
\end{tabular} \label{pb}
\end{table}

Additionally, in the reconstruction of the background components of CUORE-0 to obtain the two-neutrino DBD half-life of $^{130}$Te \cite{cuore02nu}, together with cosmogenic products in tellurium and copper, the induced $^{108m}$Ag in lead was quantified through their gamma emissions. This isotope must have been generated by neutron interactions on silver impurities present in the archaeological lead used.

\subsection{Stainless Steel}
\label{secss}

The use of stainless steel is also very common in different components of the experimental set-ups and some studies for its cosmogenic activation giving production rates of different isotopes are available both from calculations and from natural irradiation.

Activation was experimentally determined at LNGS \cite{lngsexposure}. One of the samples of stainless steel (with masses of tens of kg each) provided by Nironit company and firstly screened at LNGS using the GeMPI detector in the context of GERDA \cite{steel}, was exposed to cosmic rays outside the lab for 314~days after a cooling period underground of 327~days. The rates of production were obtained for the conditions of Gran Sasso and also at sea level, by applying the corresponding correction factor (2.4); the latter are shown in Table~\ref{ss}. Cosmogenic $^{60}$Co could not be determined due to the anthropogenic content, which is very common in steel.

Production rates have been also calculated using GEANT4 and ACTIVIA \cite{mei2016}, as presented in Table~\ref{ss}. Each code reproduces better different products, but the very high cosmogenic production of $^{7}$Be observed is predicted by none of them.

In addition, in the radiopurity study made for the XENON1T detector components, for samples of different types of stainless steel, activities of $^{54}$Mn were quantified and those of some cobalt isotopes constrained from HPGe spectrometers \cite{xenon1trpurity}.

\begin{table}
\caption{Rates of production of long-lived radioisotopes induced in stainless steel at sea level. Measurements from Ref.~\cite{lngsexposure} and different calculations are presented (see text). All numbers are expressed in atoms/kg/day.}
\centering
\begin{tabular}{@{}lcccccc@{}}
\toprule
\textbf{Isotope} & \textbf{$^{7}$Be} &  \textbf{$^{46}$Sc} & \textbf{$^{48}$V} & \textbf{$^{54}$Mn } & \textbf{$^{56}$Co} & \textbf{$^{58}$Co}   \\
\midrule
Half-life (d) \cite{toi,ddep} & 53.22(6) &  83.787(16) & 15.9735 & 312.19(3) & 77.236 &  70.85(3) \\ \hline
Measurement \cite{lngsexposure} & 389$\pm$60 & 19.0$\pm$3.5 & 34.6$\pm$3.5 &  233$\pm$26 &  20.7$\pm$3.5 & 51.8$\pm$7.8 \\
GEANT4 \cite{mei2016} & 0.05 &  8.8 &  & 230 & 16 &  90  \\
ACTIVIA \cite{mei2016} & 2.05  &  18 & &  191 &  131 & 13 \\
\bottomrule 
\end{tabular} \label{ss}
\end{table}

\subsection{Titanium}
\label{secti}

Titanium has some properties which make it an alternative option to other material in some cases. The relevant cosmogenic product is $^{46}$Sc, a $\beta^-$/$\gamma$ emitter with a transition energy of 2366.5 keV and a half-life of 83.8 days.

The radiopurity of titanium was studied in the context of LUX  \cite{luxrpurity1,luxrpurity2}, measuring for $^{46}$Sc activities from 0.2 to 23~mBq/kg in several samples and observing also different scandium radioisotopes but with a shorter half-life. A sample (with a mass of 6.7~kg), after being underground for two years, was firstly screened at the Soudan Low Background Counting Facility (SOLO), exposed to cosmic rays for six months, and screened again at SOLO, finding $(4.4\pm0.3)$~mBq/kg of $^{46}$Sc \cite{lux}, in reasonable agreement with calculations from Ref.~\cite{mei2016} using GEANT4 and ACTIVIA quoted in Table~\ref{ti}. LUX-ZEPLIN is taking into account the  activation from $^{46}$Sc \cite{lztdr}.

Titanium samples of different grades were also analyzed too in the radiopurity study made for the XENON1T detector components, even if the material was finally disregarded for the cryostat. Activities of $^{46}$Sc were quantified from HPGe measurements at the level of 1 to 3 mBq/kg \cite{xenon1trpurity}.

\begin{table}
\caption{Rates of production of long-lived radioisotopes induced in titanium at sea level. Measurements from different calculations are presented (see text). All numbers are expressed in atoms/kg/day.}
\centering
\begin{tabular}{@{}lcc@{}}
\toprule
& \textbf{$^{46}$Sc} & \textbf{$^{40}$K} \\
\midrule
Half-life \cite{ddep} & 83.787(16)~d & 1.2504(30)$\times 10^{9}$~y  \\ \hline
GEANT4 \cite{mei2016} & 275.5 &  22.1 \\
ACTIVIA \cite{mei2016}  &  270.1 &  61.0 \\
\bottomrule
\end{tabular} \label{ti}
\end{table}

\subsection{Aluminum}
\label{secal}

Aluminum is often used for metalization of germanium detector surfaces and in the cryostats of HPGe detectors or detector holders. $^{22}$Na and $^{26}$Al have been identified as the main cosmogenic products in aluminum, induced by neutron and proton reaction on $^{27}$Al. The latter, $^{26}$Al, disintegrates by electron capture and $\beta^+$ emission with a transition energy of 4004.2 keV and a half-life of 7.17$\times10^5$ y. Aluminum is usually obtained from bauxite deposits in surface mines with negligible soil overburden; and these deposits are old enough to let $^{26}$Al reach a saturation activity.

In Ref. \cite{almajorovits} aluminum was analyzed as a source of background for experiments requiring low background; the cosmogenic production by nucleons of $^{22}$Na and $^{26}$Al was computed from measured selected production cross sections and the corresponding fits and the neutron spectrum from \cite{ziegler}. Table \ref{al} shows the corresponding production rates obtained at sea level, which were evaluated also for different overburden conditions. Measured activities of $^{22}$Na and $^{26}$Al in several high purity aluminum samples were cross-checked with the expected activities from the deduced rates.

The calculations using GEANT4 and CONUS software packages and ACTIVIA to obtain production rates of different radioisotopes and to estimate of the corresponding counting rates in the Obelix HPGe detector \cite{obelix} included, together with germanium and copper, also aluminum used for the cryostat and detector holder. Table \ref{al} presents the estimated saturation activity at sea level of $^{26}$Al; production rates of $^{22}$Na in Al+4\%Si alloy were similarly computed.

From these studies, it is concluded that the contamination of $^{26}$Al can be suppressed if aluminum refined from underground deposits is used, while that of $^{22}$Na must be limited by controlling the exposure times on surface.

\begin{table}
\caption{Rates of production of long-lived radioisotopes induced in aluminum at sea level. Measurements from different calculations are presented (see text). All numbers are expressed in atoms/kg/day.}
\centering
\begin{tabular}{@{}lcc@{}}
\toprule
& \textbf{$^{22}$Na} & \textbf{$^{26}$Al} \\
\midrule
Half-life (y) \cite{ddep} & 2.6029(8) & 7.17(24)$\times10^5$  \\ \hline
Calculation for neutrons \cite{almajorovits} & 153 & 389 \\
Calculation for protons \cite{almajorovits} & 24 & 47 \\ 
ACTIVIA \cite{obelix} & & 160 \\
CONUS \cite{obelix} & & 530\\                           
\bottomrule
\end{tabular} \label{al}
\end{table}

\subsection{Argon}
\label{serar}

There is a growing need for radiopure argon for different dark matter, neutrino and DBD experiments, as liquid argon offers important advantages for radiation detection: high scintillation yield, excellent particle identification capabilities and easy purification for non-noble contaminants. Argon derived from the atmosphere contains predominately stable $^{40}$Ar, but cosmogenically produced long-lived radioactive isotopes $^{37}$Ar, $^{39}$Ar and $^{42}$Ar can be a significant background for those argon-based detectors demanding low-background. Commercial argon is produced from air and then these argon radionuclides can represent irreducible backgrounds. The worldwide low-radioactivity argon needs and the challenges associated with its production and characterization have been carefully addressed in \cite{pnnlworkshop}.

In particular, in the GERDA experiment, liquid argon acts at the same time as refrigerant and (passive and active) shielding. However, cosmogenic $^{42}$Ar present in natural argon poses a very relevant source of background. It decays with half-life 32.9 years into $^{42}$K, a $\beta^-$ emitter with transition energy of 3525 keV and with a half-life of 12.36 h; high energy electrons emitted very close to the detector surface can affect the region of interest for the neutrinoless DBD of $^{76}$Ge. Presence of $^{42}$Ar at a higher level than expected assuming a natural abundance was observed in GERDA Phase I, which was attributed to an accumulation effect due to the attraction of the generated $^{42}$K towards the germanium detectors. Different solutions have been considered; for the Phase II of GERDA, a nylon mini-shroud to screen the electric field of the detectors and create a barrier to avoid ion collection has been proved as an efficient method to reduce the events from $^{42}$K; in combination with the background rejection from Pulse Shape Discrimination in BEGe detectors and LAr veto, this background is reduced by more than a factor 1000, sufficient for Phase II of the experiment \cite{gerda42ar}. From this study, the possible level of background in the region of interest from $^{42}$K is estimated to be $<[0.14,0.5]\times10^{-3}$~counts keV$^{-1}$ kg$^{-1}$ y$^{-1}$.

There are two mechanisms for the production of $^{42}$Ar in atmospheric argon: a two-step neutron capture (requiring a high neutron flux because of the half-life of $^{41}$Ar, being of 1.8 h) and the ($\alpha$,2p) reaction on $^{40}$Ar. The specific activity of $^{42}$Ar has been studied in the context of different experiments using argon like ICARUS \cite{icarus}, DBA giving
92$^{+22}_{-46}$ $\mu$Bq/kg \cite{dba} and, more recently, DEAP, measuring 40.4$\pm$5.9 $\mu$Bq/kg \cite{deap}.

For dark matter searches, $^{37}$At and $^{39}$Ar are much more relevant. The latter is a $\beta^{-}$ emitter with a transition energy of 565~keV and half-life of 269~y; $^{39}$Ar is mainly generated by the $^{40}$Ar(n,2n)$^{39}$Ar reaction started by cosmic neutrons. The typical activity of $^{39}$Ar in atmospheric argon is around one Bq/kg, as quantified by WARP \cite{warp} and DEAP \cite{deap}. But after a campaign of extracting and purifying argon from deep CO$_2$ wells in Colorado, US, the DarkSide-50 experiment (operating a two-phase liquid argon TPC at LNGS) presented results from the use of this underground argon for the first time; the measured activity of $^{39}$Ar was (0.73$\pm$0.11) mBq/kg, which means a reduction of a factor 1400 relative to the atmospheric argon \cite{darkside}. Cosmogenically produced $^{37}$Ar was also detected in the early running of the DarkSide-50 detector, but $^{42}$Ar was not observed. The Global Argon Dark Matter Collaboration (GADMC) is working on the procurement of large amounts of radiopure underground argon developing the Urania (Colorado, US) and Aria (Sardinia, Italy) facilities for its extraction and further purification.

Understanding the production rates of cosmogenic isotopes is relevant for argon experiments, even if the use of underground argon is planned, to control their ingrowth during the different phases of the detector construction. In Ref. \cite{saldanha}, the production rates of $^{37}$Ar and $^{39}$Ar from cosmic neutrons at sea level has been measured through controlled irradiation at LANSCE with the neutron beam resembling the cosmic neutron spectrum and later direct counting with sensitive proportional counters at Pacific Northwest National Laboratory (PNNL). Results are summarized is Table \ref{ar}. In addition, the study of other production mechanisms due to protons, muons and photons was made using available cross sections to compute total production rates at sea level.

Regarding the tritium production in argon, also presented in Table \ref{ar}, production rates have been estimated using TALYS \cite{mei} and GEANT4 and ACTIVIA (with the Gordon et al. cosmic neutron spectrum) \cite{mei2016}. In  Ref.~\cite{tritiumpaper}, the study carried out and validated with experimental data for different targets (as sodium iodide and germanium) was also made for argon, from selected excitation functions and considering the Gordon et al. parameterization for neutrons. In principle, gas purification removes tritium and it should not pose a problem either for dark matter searches.

\begin{table}
\caption{Rates of production of long-lived radioisotopes induced in argon at sea level. Measurements from Ref.~\cite{saldanha} and different calculations are presented (see text). All numbers are expressed in atoms/kg/day.}
\centering
\begin{tabular}{@{}lccc@{}}
\toprule
& \textbf{$^3$H} & \textbf{$^{37}$Ar} & \textbf{$^{39}$Ar} \\
\midrule
Half-life \cite{ddep,toi} & 12.312(25)~y & 35.01(2)~d & 269~y \\ \hline
Measurement (neutrons) \cite{saldanha} & &  51.0$\pm$7.4 & 759$\pm$128 \\
Measurement+Calculations (total) \cite{saldanha} &  & 92$\pm$13 & 1048$\pm$133 \\
TENDL+HEAD \cite{tritiumpaper} & 146$\pm$31 & &  \\
TALYS \cite{mei} & 44.4 & & \\
GEANT4 \cite{mei2016} & 84.9 & & \\
ACTIVIA \cite{mei2016} & 82.9 & & \\
\bottomrule
\end{tabular} \label{ar}
\end{table}

Concerning underground activation due to muons, production rates of different nuclei induced in materials relevant for neutrino detectors, like Ar, were evaluated in Ref.~\cite{oconnel}, at a depth of 2700 m.w.e. and sea level.  
Also for argon neutrino detectors, production of different radioisotopes like $^{40}$Cl ($\beta^{-}$ emitter with a transition energy of 7482 keV generated by muon capture or (n,p) and (p,n) reactions induced by muons) was analyzed in Ref.~\cite{barker} using GEANT4 and analytic models and considering different depths.

\subsection{Other materials}

Some calculations have been made to estimate production rates in other materials commonly used in rare event experiments too. In Ref.~\cite{mei2016}, GEANT4 and ACTIVIA calculations for the production of some nuclides in PTFE are shown, as made for many other targets. ACTIVIA was also used in Ref.~\cite{pettus} to compute some production rates for quartz.

Activation studies for targets specifically used in dark matter searches have been made too.
\begin{itemize}
\item Cosmogenic isotopes have been identified in NaI(Tl) detectors used in DAMA/LIBRA \cite{naidama}, COSINE \cite{cosinebkg,pettus} and ANAIS \cite{naiepjc2016,naiepjc2019} experiments. Specific studies have been made to quantify production rates of several nuclides like some iodine and tellurium isotopes from data taken in ANAIS \cite{naijcap,naireview} and COSINE \cite{naicosine} and $^3$H and $^{22}$Na could have a relevant impact in the very low energy region.
\item Cosmogenic activation was very relevant in the scintillating bolometers made of CaWO$_{4}$ used in the CRESST-II experiment \cite{cresstbkg}, as distinct gamma lines were observed from activation of W isotopes. The rate of production of tritium in this compound was computed from TALYS \cite{mei} too.
Cosmogenic activity, as thet from other origin, was measured in Ref.~\cite{damaijmpa2} for several inorganic scintillators, including CaWO$_{4}$.
\item Silicon is a widely used detector material because it is available with very high purity and the eV-scale energy thresholds provide sensitivity to low mass dark matter particles. In addition, SiPMs are being considered for several dark matter and also DBD experiments. Tritium production is very relevant in silicon detectors; it has been computed for CDMS \cite{cdmslite} and it is considered, together with $^{32}$Si, a dominant contribution in CCDs for DAMIC \cite{damicbkg} and in DEPFET detectors \cite{depfet}. Calculations of tritium yields are also available in \cite{mei2016} and following the approach applied in Ref. \cite{tritiumpaper} for other targets. Indeed, production rates of $^3$H together with $^7$Be and $^{22}$Na have been recently obtained through controlled irradiation of silicon CCDs (from the DAMIC experiment) and wafers with the neutron beam resembling the cosmic neutron spectrum at LANSCE, followed by measurements from the CCDs in Chicago and screening of wafers with a BEGe detector at PNNL \cite{saldanhasi}. Complementing the results from the neutron irradiation with the estimates of activation for cosmic-ray particles other than neutrons (protons, photons and muons), total sea-level production rates have been derived too. Results for all production rates in silicon are summarized in Table \ref{si}.
\end{itemize}

\begin{table}
\caption{Rates of production of long-lived radioisotopes induced in silicon at sea level. Measurements from Ref.~\cite{saldanhasi} and different calculations are presented (see text). All numbers are expressed in atoms/kg/day.}
\centering
\begin{tabular}{@{}lccc@{}}
\toprule
& \textbf{$^3$H} & \textbf{$^7$Be} & \textbf{$^{22}$Na} \\
\midrule
Half-life \cite{ddep} & 12.312(25)~y & 53.22(6)~d & 2.6029(8)~y \\ \hline
Measurement (neutrons) \cite{saldanhasi} & 112$\pm$24 &  8.1$\pm$1.9 & 43.0$\pm$7.1 \\
Measurement+Calculations (total) \cite{saldanhasi} & 124$\pm$24 & 9.4$\pm$2.0 & 49.6$\pm$7.3 \\
TENDL+HEAD  as \cite{tritiumpaper} & 120$\pm$23 & &  \\
TALYS+INCL++-ABLA \cite{cdmslite} & 124 & & \\
GEANT4 \cite{mei2016} & 27.3 & & \\
ACTIVIA \cite{mei2016} & 108.7 & & \\
\bottomrule
\end{tabular} \label{si}
\end{table}

It is worth noting that, as described for argon (see Sec.\ref{serar}), in the context of experiments using large liquid scintillator detectors many results  for underground activation mainly induced by muons were obtained too for this medium, from irradiation experiments, data analysis of experiments like KamLAND and Borexino and FLUKA simulations. A summary of the works from \cite{hagner,galbiatti,kamland,borexino} was made in Ref.\cite{cebriancosmogenic}.

\section{Summary and Conclusions}
\label{secsum}

The DBD is a nuclear transition proposed to happen with extremely low probability. The two-neutrino channel has been observed for several nuclei. The neutrinoless mode, not evidenced yet, is being investigated due to the outstanding implications of its existence for the mass and other properties of neutrinos. Experiments based on different and complementary techniques, with increasing size and sophistication, are underway to identify this process. The next generation of DBD detectors require a tonne-scale size and close-to-zero  background levels with the aim to explore neutrino masses with inverted-hierarchy.

The operation in deep underground sites and the implementation of specialized background suppression methods are mandatory for DBD experiments. The long-lived radioisotopes induced in materials due to the exposure to cosmic nucleons on the surface of the Earth can be an issue for all experiments requiring ultra-low levels of the radioactive background; even the production of short-lived nuclei by underground muons must be taken under control in some projects. As the primordial activity of materials is reduced and the experiment sensitivity is larger, cosmogenics are becoming increasingly relevant. Production rates and induced activities of many different cosmogenic radionuclides in different materials (making up the detector medium or used in secondary systems) have been estimated from measurements or calculations in the frame of experiments investigating dark matter and neutrinos, including those for DBD.

Activation in germanium has been extensively studied for decades, from many different approaches. Experiments investigating low probability events using germanium detectors have derived in the last years from the analysis of their data very important results for quantifying the yields of several isotopes. Sea-level rates of production in enriched germanium used in DBD experiments are typically lower than in natural germanium. Unfortunately, there is not a precise quantification of the products relevant for the neutrinoless DBD of $^{76}$Ge, $^{60}$Co and $^{68}$Ge; the identification of $^{60}$Co in the detector medium is hard due to the simultaneous detection of beta and gamma emissions and estimates of $^{68}$Ge show an important dispersion of results. Shiedings to cosmic rays for storage and transport have been implemented for germanium detectors and even underground germanium crystal growth and detector fabrication has been sometimes proposed. Underground production by muons of $^{77m}$Ge has been also analyzed and seems to be under control for future germanium DBD projects. Activation in tellurium has been also deeply studied, from calculations and irradiation experiments made on purpose. $^{60}$Co and $^{110m}$Ag were identified as the most relevant products in tellurite crystals, but their effect on the measured background levels is not relevant thanks to the limited exposure to cosmic rays. For xenon, the result dispersion for most products is significant but spallation products identified are not relevant for the neutrinoless DBD search of $^{136}$Xe; however, $^{137}$Xe, produced by neutron capture on the DBD emitter, has been identified and analyzed as a significant background in xenon DBD experiments. Activation in other materials used in the set-up, like copper components close to the detector or argon used as cooling medium and veto, has been also carefully analyzed as some products could give a non-negligible contribution if not taken into consideration.

Precise quantification of cosmogenics is difficult as there are limited experimental data available for validation and theoretical predictions based on different approaches suffer in many cases from a large spread. The Gordon et al. parameterization \cite{gordon} for the cosmic neutron spectrum is mostly used in recent cosmogenic studies, but a good knowledge of the variation in time and from place to place of the different components of the cosmic ray flux is necessary to make reliable estimates. Uncertainties come also from the difficulties encountered on the evaluation of excitation functions at all energies; below some hundreds of MeV, where the differences for proton and neutron cross sections are the most relevant, few neutron data are available.

A detailed knowledge of the material history concerning the exposure to cosmic rays is necessary to derive from measurements production rates. Important results for some target materials have been obtained in the last years either from dedicated experiments exposing materials to beams or directly to cosmic rays in well-controlled situations or from the careful analysis of the data taken in rare event experiments. They are very relevant not only because production rates are essential to quantify cosmogenic activities for experiments but also because these measurements help to validate models and codes used in calculations.

All in all, it can be considered that cosmogenic activation is not usually the major background component for DBD experiments, provided it is adequately taken into consideration by reducing as much as possible exposure on surface during production and transport of components or by applying material purification techniques. For this reason, tools capable of reliably assessing cosmogenic yields are highly recommended and a great effort is being devoted to have more accurate estimates of the impact of cosmogenic activation for experiments searching for DBD and other rare phenomena.






\conflictsofinterest{The author declares no conflict of interest.}


\reftitle{References}




\end{document}